%% file: article.tex
\documentclass[12pt]{article}
\usepackage{epsfig}
\usepackage{xspace}
\input econfmacros.tex
\def\Title#1{\begin{center} {\Large {\bf #1} } \end{center}}
\newcommand*{\ttbar}{\ensuremath{t \bar{t}}\xspace}
\newcommand*{\TeV}{\ifmmode {\mathrm{\ Te\kern -0.1em V}}\else
                   \textrm{Te\kern -0.1em V}\fi}%
\newcommand*{\GeV}{\ifmmode {\mathrm{\ Ge\kern -0.1em V}}\else
                   \textrm{Ge\kern -0.1em V}\fi}%
\newcommand*{\MeV}{\ifmmode {\mathrm{\ Me\kern -0.1em V}}\else
                   \textrm{Me\kern -0.1em V}\fi}%
\let\tev=\TeV
\let\gev=\GeV
\let\mev=\MeV
\def\iab{\mbox{ab$^{-1}$}}
\def\ifb{\mbox{fb$^{-1}$}}

\begin{document}

\Title{Top physics at high-energy lepton colliders}

\bigskip\bigskip


\begin{raggedright}  

{\it Summary of TopLC15, IFIC Valencia, 30$^{th}$ June - 2$^{nd}$ July, 2015\\
\bigskip
M.~Vos (IFIC, editor) \\
Attendants of the workshop: \\ 
G.~Abbas (IFIC), M.~Beneke (TUM), S.~Bilokin (LAL), M.J.~Costa (IFIC), S.~de~Curtis (U. \& INFN Firenze), K.~Fujii (KEK), J.~Fuster (IFIC), I.~Garcia~Garcia (IFIC), P.~Gomis (IFIC), A.~Hoang (U. Vienna), A.~Irles (DESY), Y.~Kiyo (Yuntendo), M.~Kurata (Tokyo), L.~Linssen (CERN), J.~List (DESY), M.~Nebot (Lisboa), M.~Perello (IFIC), R.~P\"oschl (LAL), N.~Quach (KEK), J.~Reuter (DESY), F.~Richard (LAL), G.~Rodrigo (IFIC), Ph.~Roloff (CERN), E.~Ros (IFIC), F.~Simon (MPI Munich), J.~Tian (KEK), A.F.~\.Zarnecki (Univ. of Warsaw) \\
\bigskip
Corresponding author: M. Vos, IFIC (UVEG/CSIC), Edificios de Investigacion, \mbox{c./ Catedratico} Jose \mbox{Beltran 2}, E-46980 Paterna, Valencia, SPAIN (marcel.vos@ific.uv.es)}
\bigskip\bigskip
\end{raggedright}

\begin{abstract}
A summary is presented of the workshop ``top physics at linear colliders'' that was held at IFIC Valencia from the 30$^\mathrm{th}$ of June to the 3$^\mathrm{nd}$ July 2015. We present an up-to-date status report of studies into the potential for top quark physics of lepton colliders with an energy reach that exceeds the top quark pair production threshold, with a focus on the linear collider projects ILC and CLIC. This summary shows that such projects can offer very competitive determinations of top quark properties (mass, width) and its interactions with other Standard Model particles, in particular electroweak gauge bosons and the Higgs boson. In both areas the prospects exceed the LHC potential significantly - often by an order of magnitude. 
\end{abstract}

\section{Introduction}

Whereas hadron colliders have dominated the landscape of high-energy 
particle physics for well over a decade, high-energy lepton colliders 
feature prominently on the roadmap for the future of particle physics. 
Their precision physics programme
forms an ideal complement to the discovery reach of the LHC. 

A mature, shovel-ready project exists for a linear $e^+e^-$ collider that
can reach a center-of-mass energies from several 100~\gev{} to approximately 
1~\tev{} (the International Linear Collider or ILC~\cite{Baer:2013cma}, 
to be hosted at the Kitakami site in Japan). Extensive R\&D into high-gradient 
acceleration has moreover opened up the possibility of a relatively compact 
multi-\tev{} collider (Compact Linear Collider, or CLIC~\cite{Linssen:2012hp}). 
More recently, renewed attention has been devoted to the possibility of a 
large circular $e^+e^-$ collider, the triple-LEP~\cite{Gomez-Ceballos:2013zzn}.
An $e^+e^-$ collider with center-of-mass energy up to the \ttbar{} threshold
could form the first stage of the Future Circular Collider (FCC) at CERN. 
In China an only slightly less ambitious project is being 
pursued~\cite{CEPC-SPPCStudyGroup:2015csa}, that could host a 
250~\gev{} $e^+e^-$ collider (CEPC) in its
initial stage. A muon collider~\cite{Alexahin:2013ojp} is explored for a more 
remote future. In this report
the focus is on the linear collider projects, for which detailed
experimental studies have been performed. Quite often, however, conclusions 
apply to $e^+e^-$ colliders in general. Wherever relevant studies
are available from circular machines, we will include them in the discussion.

The case for a high-energy lepton collider rests strongly on the potential
to characterize the couplings of the Higgs boson discovered in 2012 to
sub-percent precision~\cite{Dawson:2013bba}. Also the potential of lepton 
colliders to open a complementary window on new physics (from leptophilic
dark matter to sleptons) is well established~\cite{Gershtein:2013iqa}. 
The third pillar of the programme is a detailed scrutiny of the top 
quark~\cite{Agashe:2013hma}. 

The TopLC15 workshop in Valencia is the third in a series of workshops 
devoted to top physics at future lepton colliders~\footnote{Previous
editions were organized in the greater Paris area, by LAL Orsay and 
LPNHE. The fourth 
workshop will be held at KEK in Tsukuba, Japan, from 6-8 July 2016.}. 
The aim of the series is to enhance the cohesion of the global effort to 
understand the top physics potential of lepton collider fully. In particular, 
we hope the TopLC workshops bring together theorists and experimentalists.
This summary report aims to provide an up-to-date reference and 
bibliography. 

The focus of the workshop was on several measurements of top quark 
properties and interactions where detailed
prospect studies exist, such as the mass measurement at the pair
production threshold and the measurements of the top quark couplings 
to the Higgs boson~\cite{Price:2014oca} and
the photon and $Z$-boson~\cite{Amjad:2015mma}. Sections~\ref{sec:mass},
\ref{sec:tth} and~\ref{sec:ttz} present an up-to-date status report
for these studies.  
In Section~\ref{sec:fcnc} we enter a less explored area, discussing
recent parton-level studies of the potential of the linear collider 
projects to detect 
Flavour Changing Neutral Current (FCNC) decays of the top quark.
The impact of reconstruction algorithms on the top quark physics potential 
is discussed in Section~\ref{sec:reco}, with a focus on the 
reconstruction of tracks and jets.
The final Section~\ref{sec:summary} presents a brief summary and outlook.

\section{Top quark mass}
\label{sec:mass}
The top quark mass is one of the key parameters of the Standard Model of 
particle physics. A precise determination enables stringent tests of the
self-consistency of the theory. A precise measurement is needed to verify
the relation between the top quark mass and the Higgs boson and 
W-boson masses predicted by the SM (see, for instance, Ref.~\cite{Baak:2014ora}).
The value of the top quark mass moreover has a strong impact 
on the stability of the vacuum when the Higgs potential is extrapolated to large
scales~\cite{Degrassi:2012ry}. In the following we present
the current precision and the prospects of the complete LHC programme.
We also discuss the uncertainties that affect the interpretation of 
the {\em direct} measurement of the top quark mass and alternatives pursued
by the LHC experiments.
We then present the ultimate precision achievable at a lepton collider 
that scans its center-of-mass energy through the top quark pair production 
threshold region. After a brief presentation of two recent theory milestones, 
the calculation of the NNNLO correction to the cross-section at threshold
 and the four-loop relation between different mass schemes, we discuss
the remaining theory uncertainty. We finalize this Section with the 
most up-to-date prospects of the linear collider for the top quark mass 
determination, including realistic estimates for the theoretical and 
experimental systematic uncertainties.

\subsection{LHC, state of the art and prospects} 

The current world average for the top quark mass based on {\em direct} 
measurements at hadron colliders have attained a precision of better 
than 0.5\%~\cite{ATLAS:2014wva}. The most precise measurements by CMS 
and D0 achieve 500~\mev{} uncertainty per measurement. 
After three years of operation
at approximately half the design energy the ATLAS and CMS already exceed the 
expectations~\cite{Beneke:2000hk} drawn up before the start of the LHC. 

With the large \ttbar{} samples collected at the LHC the statistical
uncertainty on the measurement ceases to be relevant.
The current measurements at the LHC are already dominated by 
systematic uncertainties, with the most important contributions coming
from the uncertainties on the jet energy scale and 
in the modelling of the \ttbar{} signal. It is therefore far from
straightforward to draw up reliable prospects for the future
evolution of the uncertainty. Expectations range from a pessimistic
500~\mev{}~\cite{Juste:2013dsa} after the complete LHC programme
to 200~\mev{}~\cite{CMS-PAS-FTR-13-017}. These estimates 
explicitly exclude the theoretical uncertainty from 
the total error budget. An additional uncertainty must be added to account 
for the ambiguity in the interpretation of the top mass parameter 
in the Monte Carlo generators in terms of a rigorous 
field-theoretical mass scheme, as discussed in the next section.

Alternative mass determinations at hadron colliders include a determination 
of the top quark mass from a fully corrected cross-section measurement.
The cross-section measurements by ATLAS and CMS in the di-lepton ($e\mu$) 
channel on the 7 and 8~\tev{} data~\cite{CMS-PAS-TOP-13-004,Aad:2014kva} 
reach a precision of approximately 4\% and have a negligible dependence
on the (MC) mass assumed in the correction of the acceptance.
The pole mass is extracted by comparing the observed cross-section to 
the NNLO calculation of Ref.~\cite{Czakon:2013goa} with NNLL resummation, 
which reduces the scale uncertainty on the
cross-section to the level of 3\%. An uncertainty of 1.7$-$1.8\% is assigned
to the cross-section to account for the uncertainty in the LHC beam energy.
The value of the top quark pole mass that is determined 
is in agreement within the uncertainty of approximately 2~\gev{} with the 
result of the {\em direct} mass measurement.
Further progress is expected from improved PDF fits.

Adrian Irles presented the result of a new pole mass measurement on 7~\tev{}
LHC data by 
ATLAS~\cite{Aad:2015waa}, where the mass is extracted
from the differential cross-section in top quark pair production in association
with a hard jet, following the method proposed in Ref.~\cite{Alioli:2013mxa}.  
The result is again in good agreement (within an uncertainty of 2.3~\gev{})
with the other determinations. As this measurement is not limited by the
systematic uncertainty of relating the MC mass to the top quark mass,
an analysis of the 8~\tev{} and 13~\tev{} data
sets can bring a strong improvement of the precision.

\subsection{Top quark mass, theory and interpretation}

The most precise measurements of the top quark mass at hadron colliders extract
the top quark mass by comparing distributions generated using Monte Carlo (MC) 
generators to the data. In the standard interpretation the 
MC mass parameter is identified with the pole mass. This interpretation
has an ambiguity that is estimated to be ${\cal O} (1\gev)$~\cite{Juste:2013dsa,ahoang08,Moch:2014tta,ahoang14,Corcella:2015kth}.
The uncertainties cited by the experiments include contributions
for the modelling of $\ttbar{}$ production and decay, non-perturbative
corrections, colour reconnection, etc. At the precision of today's 
measurements these may adequately cover the intricacies in the standard 
interpretation of the top quark mass, but for progress towards a 
200~\mev{} top quark
mass measurement a more sophisticated treatment seems required.

Several theory groups are performing studies to elucidate the relation between
the MC mass parameter and field theory mass definitions~\cite{ahoang14,Corcella:2015kth}.
At the workshop Andre Hoang showed preliminary results from a study 
that compare the predictions of mainstream Monte Carlo generators
(such as Pythia~\cite{Sjostrand:2006za}) to hadron-level QCD calculations.
The latter are based on the work in Refs.~\cite{Gritschacher:2013pha,Gritschacher:2013tza,Pietrulewicz:2014qza} in Soft Collinear Effective field Theory (SCET) and account for perturbative and non-perturbative effects. A comparison of both predictions yields a relation
between the mass parameter of the Monte Carlo generator and the
quark mass in the calculation.
The distributions under study include the thrust in 
$e^+ e^- \rightarrow b \bar{b} $ and $e^+ e^- \rightarrow t \bar{t}$ 
pair production at different center-of-mass energies. These preliminary
results indicate that the relation between MC mass and field theoretical
top quark mass definitions
may be established to a precision of 500~\mev{} for bottom quarks
and better than 1~\gev{} for top quarks.

Alternative measurements, such as those discussed in the previous section,
form an independent cross-check of the interpretation of the {\em direct} 
measurement, provided they can achieve sub-\gev{} precision.

\subsection{The $e^+ e^- \rightarrow t \bar{t}$ production threshold: theory status}

The top quark pair production threshold at electron-positron colliders
has been identified long ago~\cite{Gusken:1985nf} as a key element in the 
programme of high-energy lepton colliders. The position of the threshold 
is related to the top quark mass $m_t$, the slope of the rise in 
cross section around the threshold reflects its natural width. 
The cross section in the threshold region is moreover sensitive 
to the top quark Yukawa coupling and the strong coupling constant $\alpha_s$. 
A precise measurement of the shape
of the sharp rise of the cross-section around $\sqrt{s} = 2 m_{t}$ 
can provide competitive measurements of these parameters.

To take advantage of the potential of a threshold scan precise
predictions of the threshold shape are crucial. As QCD bound-state effects
become sizeable, the calculations must be organized as a combined 
perturbative series in terms of the top quark velocity and $\alpha_s$. 
The state-of-the-art fixed-order NNNLO 
calculation~\cite{Beneke:2015kwa}
presented at the workshop by Martin Beneke and Yuichiro Kiyo achieve
a precision of approximately 3\% on the cross section.
More importantly, when using the PS or 1S mass schemes, the peak position 
shows excellent convergence: the NNNLO correction represents a 65~\mev{} 
shift. Higgs boson exchange~\cite{Beneke:2015lwa}, the known non-resonant
contribution at NLO~\cite{Beneke:2010mp} and electromagnetic corrections
are included in the NNNLO description, but not yet the known electro-weak 
corrections.

\begin{figure}[htb]
\begin{center}
\includegraphics[width=0.47\linewidth]{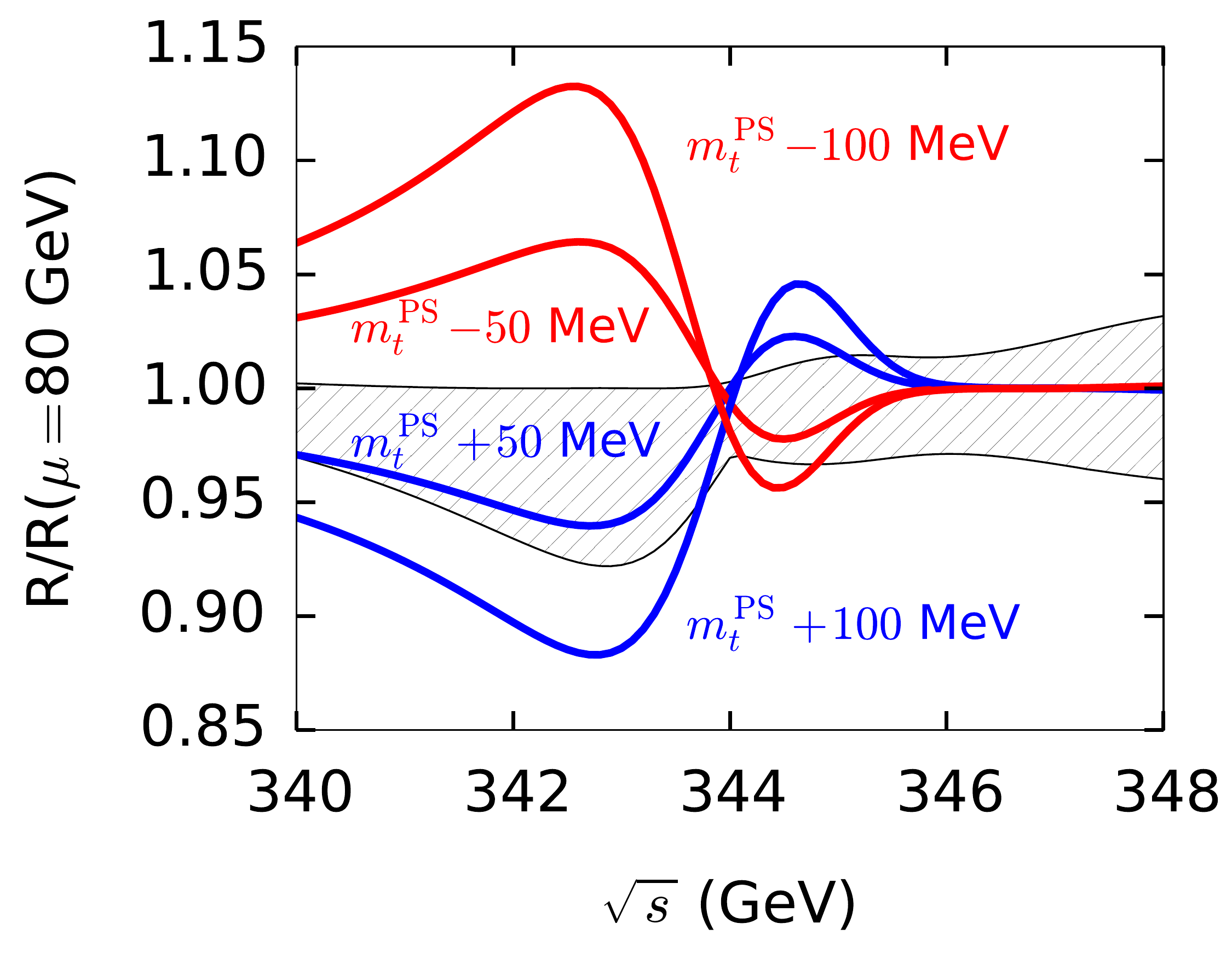}
\includegraphics[width=0.47\linewidth]{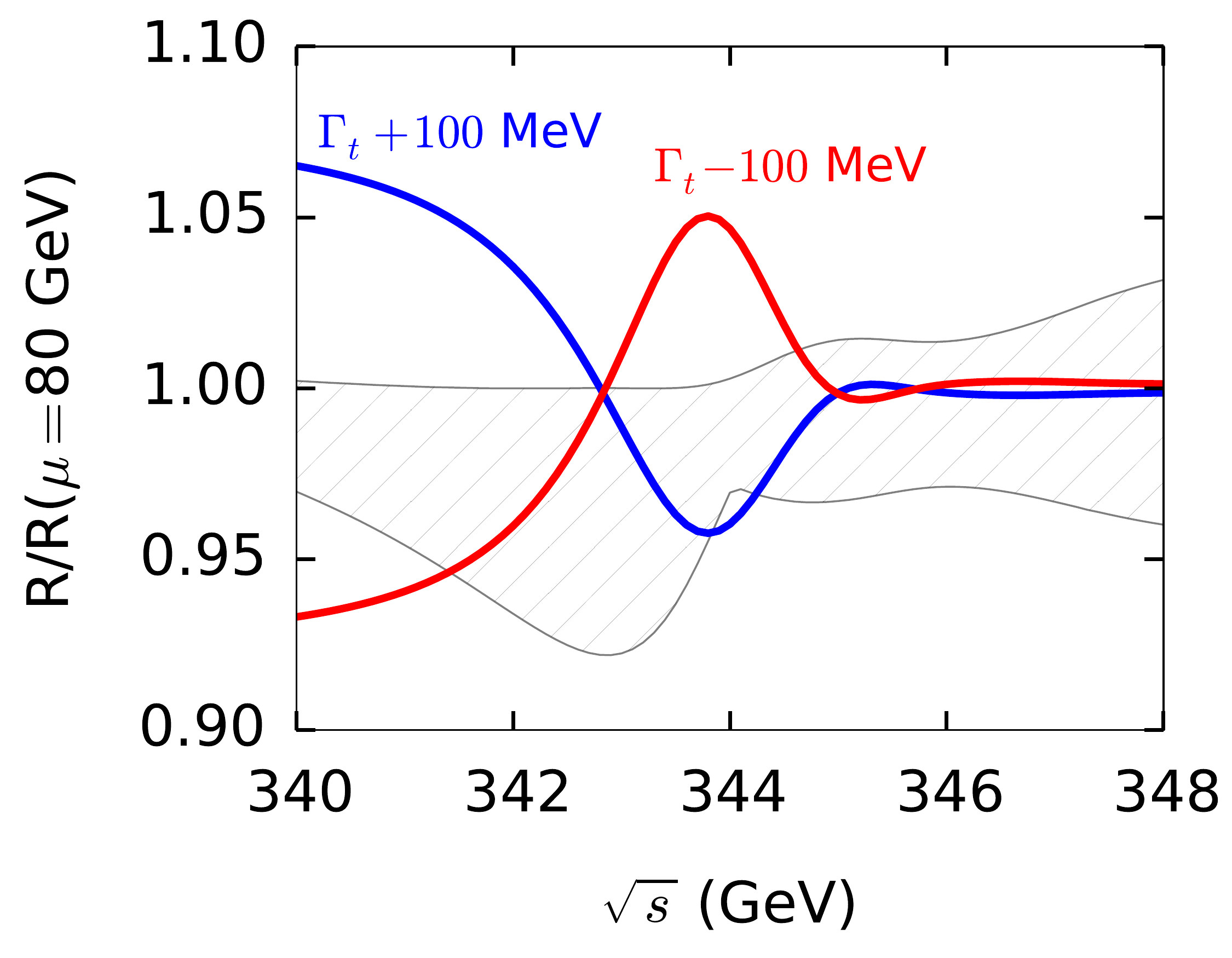}
\caption{The uncerainty on the \ttbar{} threshold shape at NNNLO precision from Ref.~\cite{Beneke:2015kwa}. In the leftmost plot the change in the shape due to a shift of the top quark mass is superposed, in the rightmost plot the effect of a change in the top quark width. }
\label{fig:threshold_uncertainty}
\end{center}
\end{figure}

The uncertainty band on the NNNLO calculation is presented in 
Fig.~\ref{fig:threshold_uncertainty}.
From a comparison of the width of the error band with the variation 
of the top quark mass by 50~\mev{} one can estimate that the systematic
uncertainty on the top quark due to this uncertainty is of the order
of tens of \mev{} and likely smaller than 50~\mev{}\footnote{In agreement
with the estimate based on the NNLL renormalization group improved 
calculations of Ref.~\cite{Hoang:2013uda}.}.
A rigorous propagation of the uncertainty in a realistic fit is
to be performed in the near future.

In Ref.~\cite{Hoang:2013uda} Hoang and Stahlhofen have performed a renormalization group improved next-to-next-to leading logarithmic (NNLL) calculation of the threshold shape in the framework of velocity Non-relativistic QCD (vNQCD) which
also accounts for the resummation of logarithms of the top quark velocity. 
A combination of the NNLL renormalization-group improved results with the 
NNNLO fixed-order calculation
has the potential to further reduce the theoretical uncertainty.

A description of the QCD effects at the pair production threshold 
with NLO accuracy is included in recent versions of the matrix element 
generator WHIZARD~\cite{Kilian:2007gr}. 
The model {\em tt threshold} 
allows for the generation of fully differential distributions.
In the $\sqrt{s}$ region immediately above the threshold the bound-state 
QCD effects gradually die out. For a reliable estimation of the cross-section
in this region the threshold calculation must be matched to the continuum 
predictions. Preliminary results from this effort were presented in 
the contribution by J\"urgen Reuter.

\subsection{Lepton collider prospects}

A threshold scan, a scan of the center-of-machine energy of the machine 
to map out the top quark pair production threshold, is part of the 
programme of all electron-positron collider projects with
sufficient energy reach. Typically, a ten-point scan is envisaged
in a narrow (10~\gev{}) region around the position of the would-be
1S resonance. Mostly, equidistant points are assumed with 
approximately 10~\ifb{} per scan point. A threshold scan with
these characteristics can be performed in a fraction of a year.

The shape of the threshold region is affected by several effects, such
as Initial State Radiation (which is equal for all machines) and the
beam energy spread (where each machine has its own characteristic profile).
The resulting threshold shapes of three $e^+e^-$ projects are shown in 
Fig.~\ref{fig:threshold_isr_beam}. Even if the 1S peak is washed out to
different degrees, a one-parameter fit of the top mass to the 
threshold shape (ten points, 10~\ifb each. no polarization) yields 
a quite similar statistical accuracy of approximately 20~\mev{} for
all projects~\cite{Simon:2016htt}.
These difference are small compared to several of the systematic 
uncertainties evaluated in the next Section, such that the specific
profile of each machine has a negligible impact on the precision
that can be reached.

\begin{figure}[htb]
\begin{center}
\includegraphics[width=0.6\linewidth]{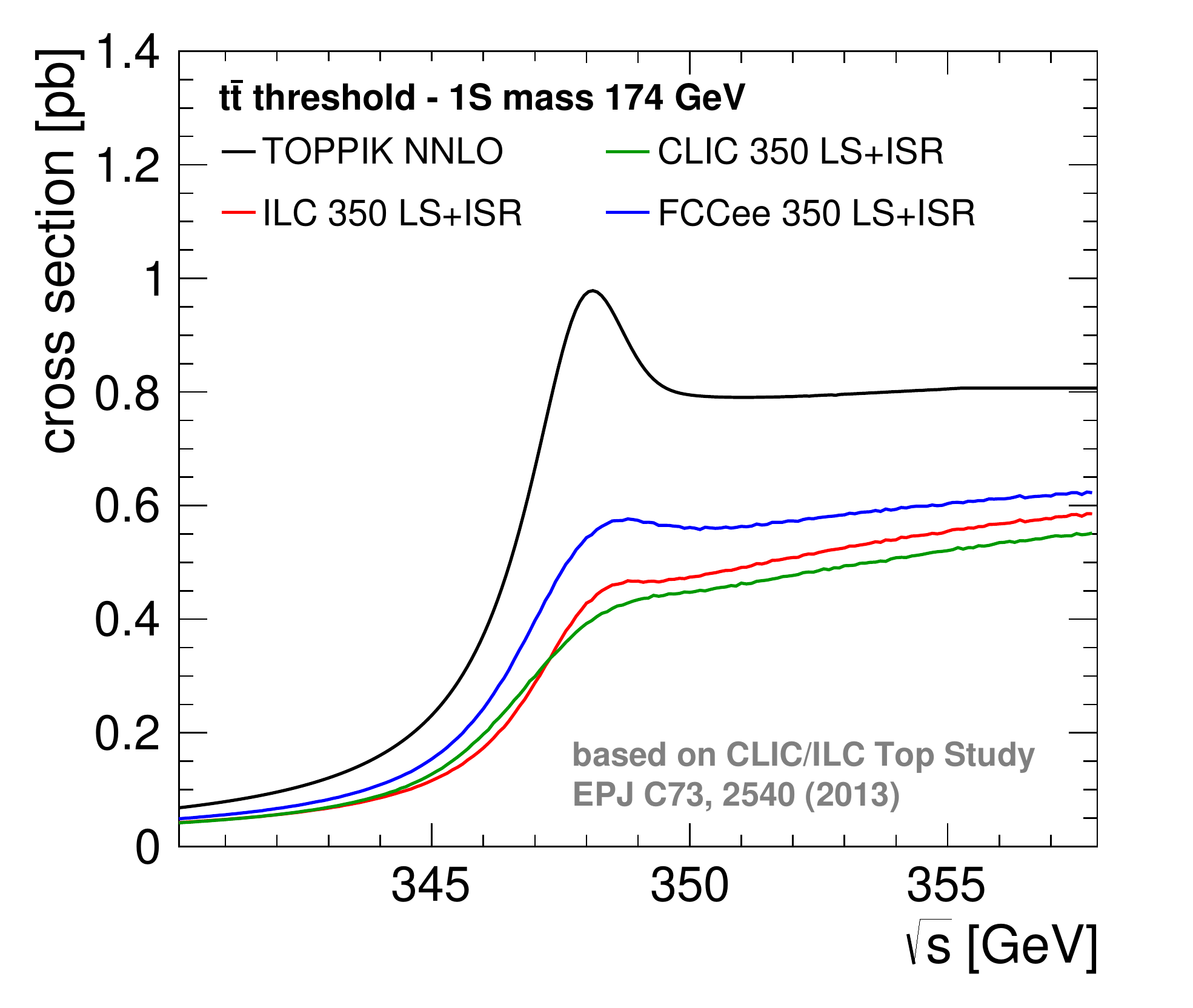}
\caption{Realistic \ttbar{} threshold shapes for several $e^+e^-$ collider projects. The theory curve from the NNLO calculation available in the TOPPIK code is folded with the luminosity spectrum (LS), which includes an estimate of the beam energy spread and the effect of beamstrahlung in the linear collider projects, and Initial-State-Radiation (ISR). }
\label{fig:threshold_isr_beam}
\end{center}
\end{figure}

The statistical uncertainty on the mass depends on the number of fit points and
on the number of floating parameters. An optimization of the location of the fit points can save considerable running time. In a one-parameter fit of the threshold shape one achieves identical precision on the top quark mass with three wisely chosen points (i.e. 30~\ifb{}) as with the full ten-point scan. The situation is more complex as soon as more than one parameter is floated. The uncertainty increases by a factor three when, apart from the mass, the top quark width and the strong coupling constant $\alpha_s$ are extracted.

\subsection{Systematic uncertainties}

Several groups have evaluated systematic uncertainties on the top quark mass
extraction. A number of experimental sources of uncertainty, including the impact of non-\ttbar{} background, are evaluated in Ref.~\cite{Seidel:2013sqa}. The largest contribution is expected to stem from the uncertainty in the beam energy, where a residual uncertainty of 1 part in 10.000 yields a 30~\mev{} uncertainty in the top quark mass. The uncertainty due to an imperfect knowledge of the luminosity spectrum was revised based on the work of Ref.~\cite{Poss:2013oea} and is expected to be of the order of 10~\mev~\cite{Simon:2014hna}. Ref.~\cite{Beneke:2010mp} has shown that the non-resonant contribution is approximately constant: its impact can be reduced to less than 30~\mev{} with a simple correction~\cite{Fuster:2015jva}. While a complete and systematic evaluation of the systematics is still missing, this patchwork of results yields an experimental systematic uncertainty of the order of 40~\mev.

The theory uncertainty on the top quark mass has two main contributions. The theory uncertainty in the NNNLO description of the threshold shape discussed in the previous Section (see Fig.~\ref{fig:threshold_isr_beam}) is expected to be below 50~\mev{}. A full-fledged fit that includes the uncertainty band has been performed since the workshop and indeed yields a theory uncertainty of approximately 45~\mev~\cite{Simon:2016htt}. An additional uncertainty in the conversion of the threshold mass to the $\bar{MS}$ scheme is discussed in the next Section. 

\subsection{Conversion to the $\bar{MS}$ mass}

Another highlight of the theory effort is the calculation by Marquard 
et al.~\cite{Marquard:2015qpa} of the conversion between different mass schemes 
to four loops. The translation of the threshold (1S or PS) mass obtained from 
the fit to the threshold shape to a short distance mass such as the 
$\bar{MS}$ mass is an intrinsic part of the measurement procedure.
In the three-loop conversion the theory uncertainty in this step 
was in fact the dominant uncertainty on the mass (approximately 100~\mev). 
With the addition of the fourth loop the scale uncertainty is reduced
to the level of 10~\mev. A non-negligible parametric uncertainty
due to the uncertainty of the strong coupling constant remains, however.

The uncertainty in the current (2014) world average~\cite{Agashe:2014kda} for
$\alpha_s$ at the Z boson mass is 0.0006. Today, the uncertainty in the conversion
therefore amounts to approximately 40~\mev{}. In a somewhat unusual
turn of events the $\alpha_s$ uncertainty is expected to increase by a factor 
two~\cite{d'Enterria:2015toz} in the next world average. 
The potential for future improvement
is somewhat uncertain~\cite{Gomez-Ceballos:2013zzn,Aoki:2013ldr}. 
For a detailed discussion the reader is referred to 
Ref.~\cite{d'Enterria:2015toz} and references therein. 

The strong coupling constant affects the extraction of the top quark $\bar{MS}$
mass from the threshold in the standard scheme in two ways: $\alpha_s$ plays
a role in the prediction of the line shape that is used to extract a
threshold mass ($1S$ or $PS$ mass, a larger value of $\alpha_s$ leads to 
a larger prediction for the cross section) and in the conversion of the 
threshold mass to the $\bar{MS}$ mass. The threshold shape 
itself provides a measurement of $\alpha_s$, but that constraint is 
somewhat weaker than the world average. 
In Figure~\ref{fig:massalphas} the precision on the
$\bar{MS}$ mass $\overline{m}_t (\overline{m}_t)$ is shown as a function 
of the uncertainty of the $\alpha_s$ prior. The contributions from 
the uncertainty on the line shape and the parameteric uncertainty
in the mass conversion are added in quadrature. 
A coherent use of $\alpha_s$ in the mass fit and conversion 
leads to a partial cancellation between both terms, such that
the curve in Figure~\ref{fig:massalphas} should be considered as
conservative upper limit.

At the workshop Yuichiro Kiyo showed that the shifts in the peak 
position due to scale variations are reduced when the NNNLO calculation
is performed in the $\bar{MS}$ scheme than with the PS mass. This
demonstrates that indeed some cancellations occur. A direct extraction
of the $\bar{MS}$ mass may then be interesting~\cite{Kiyo:2015ooa}. A detailed
study of both schemes in a realistic environment, and with a
coherent treatment of the $\alpha_s$ uncertainty, is encouraged.

\begin{figure}[htb]
\begin{center}
\includegraphics[width=0.9\linewidth]{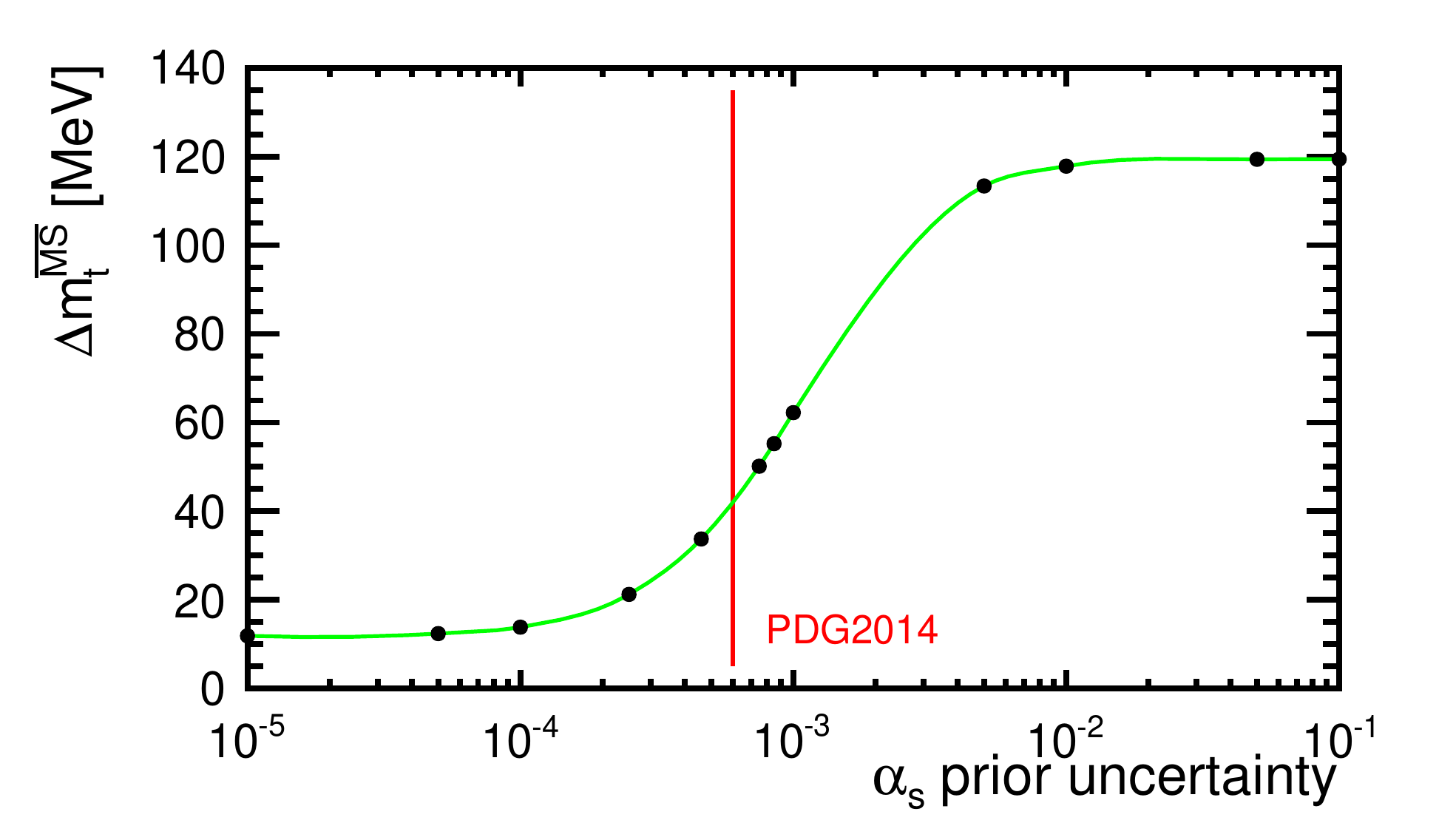}
\caption{The uncertainty on the $\bar{MS}$ mass of the top quark obtained from a threshold scan as a function of the uncertainty on the value of the strong coupling constant $\alpha_s$ that is used as a prior in the fit and the conversion of the threshold mass to a short-distance mass. Courtesy of M. Perello and M. Vos. }
\label{fig:massalphas}
\end{center}
\end{figure}

\subsection{Mass measurement in the continuum}

Mass measurements in continuum \ttbar{} production at center-of-mass
energies above the threshold are interesting for a number of reasons.
First, the threshold scan is not scheduled as the first step
of ILC or CLIC programme. Even if the measurement in the continuum
is ultimately superseded by the threshold scan it may provide
the best top quark mass at the time. The measurement of the 
top quark mass at a scale different from the \ttbar{} threshold may
provide access to the running of the top quark mass.

The full-simulation study of Ref.~\cite{Seidel:2013sqa} establishes
the statistical precision of the top quark mass measurement using the
{\em direct} measurement in the ILC operated at 500~\gev{} with
an integrated luminosity of 500~\ifb.
As the mass extraction is based on Monte Carlo templates, as it is
at the LHC, the interpretation in a rigorously defined mass scheme
remains problematic. By the same token, a precise measurement at 
a lepton collider may prove to be a valuable tool to elucidate the
interpretation of this top quark mass measurement.

At the workshop an alternative method was presented by Marca Boronat
and Pablo Gomis, that extracts the top quark mass from the $s'$ 
($s' = s (1- 2 E_{\gamma}/\sqrt{s})$) distribution 
in \ttbar{} $+ \gamma $ or \ttbar{} $+$ jet production.
The rate of energetic ISR photons (and FSR gluons) 
depends strongly on the top quark mass. A comparison of a precise
measurement of the differential cross-section to a fixed-order
calculation allows for a mass extraction with a rigorous interpretation 
in the mass scheme of choice. 
Preliminary, parton-level results indicate that a precision of 100~\mev{}
may be feasible with the \ttbar{} $+ \gamma $ analysis using 
500~\ifb{} at 380~\gev{}. At higher energy larger samples are
required: the same precision is reached only after accumulating 
4000~\ifb{} at 500~\gev{}. The statistical uncertainty of the  
\ttbar{} $+$ jet analysis is better than this, but a detailed
study of the identification of the additional jet is required.

At lepton colliders with a center-of-mass energy beyond 500~\gev{} 
the top quark mass can be extracted from the shape of boosted top quark jets,
as proposed in Refs.~\cite{Fleming:2007qr,Fleming:2007xt}.

\subsection{Summary}

The possibility of lepton colliders to scan their center-of-mass energy
through the \ttbar{} production threshold is one of the most exciting
prospects of such a machine. The threshold scan allows for a
top quark mass measurement with a statistical uncertainty of order 10~\mev{}
and a total uncertainty below 50~\mev{}. Further progress
in theory and $\alpha_s$ may improve these prospects.
Measurements of the mass using data sets acquired at center-of-mass
energies well above the threshold can reach a statistical precision of better
than 100~\mev. New methods that maintain a clean interpretation are
being developed.


\section{Top quark couplings to the Higgs boson}
\label{sec:tth}
The top quark and the Higgs boson form a {\em dream couple} for precision
measurements. The heaviest particle in the Standard Model, the top
quark, is tightly coupled to the Higgs boson with a Yukawa coupling of
the order of unity.
As the top quark is one of the main drivers of the instability
of the Higgs boson mass it is quite conceivable that the interaction
between the top quark and the Higgs boson is intimately connected to 
new physics. Many extensions of the Standard Model indeed reserve
a special role for this couple, in the form of top quark partners, 
a composite top and Higgs sector, etc. A direct measurement of the
coupling of these two particles is therefore desirable.

\subsection{LHC status and prospects}

Gluon fusion Higgs production at the Large Hadron Collider proceeds 
primarily through a top quark loop. The same is true for the decay to
a photon, one of the most prominent channels behind the discovery
of the Higgs boson in 2012. The ATLAS and CMS experiments therefore
have good indirect sensitivity to the top quark coupling to the
Higgs boson. In the 7-parameter fit that is currently the de facto
standard, the couplings to up- and down-type quark are allowed
to vary independently, but the couplings of the quarks of
a given type (i.e., down-type quarks d,s and b, or up-type quarks u,c and t)
are assumed to be identical. These fits achieve a precision of 
approximately 20\%, which
is expected to improve to the level of 7-10\%~\cite{Dawson:2013bba}.

A more direct probe of the top quark Yukawa coupling is found in the
associated production of a top quark pair and a Higgs boson. The
production cross section is nearly 1 pb at $\sqrt{s} =$ 13~\tev{},
significantly smaller than top quark pair production (nearly 1 nb)
and other Higgs boson production processes (up to 100 pb).
After run I of the LHC the ATLAS and CMS experiment have reported
first evidence for $t\bar{t}H$ production. Combining all decay channels
and the two experiments, the significance is approaching the 5$\sigma$
threshold. The best fit signal strength 
$\mu_{ttH} = \sigma/\sigma_{SM} =$ 2.3$^{+0.7}_{-0.8}$ of the ATLAS + CMS combination
in Ref.~\cite{ATLAS-CONF-2015-044} is somewhat larger 
than in the Standard Model. 

In a complex final state such as $t\bar{t}H$, with large systematics
associated to the modelling of the background, a prediction of the
LHC potential is prone to large uncertainties.
The Snowmass Higgs report~\cite{Dawson:2013bba} expects the direct 
measurement of the 
top Yukawa coupling in $t\bar{t}H$ production at the LHC to reach a precision of
14-15\% after 300~\ifb{} at $\sqrt{s}=$ 14~\tev. The full LHC programme, 
including the luminosity upgrade, could reduce this to 7-10\% (depending
on the assumptions on the evolution of systematic uncertainties) after
accumulating 3000~\ifb. 

\subsection{Lepton Collider prospects}

At electron-positron colliders $t\bar{t}H$ production proceeds through
the s-channel, with the Higgs boson radiated off one of the top
quarks (i.e. $e^+e^- \rightarrow Z/\gamma^* \rightarrow t\bar{t}H$).
The cross section of Figure~\ref{fig:tth_xsec} 
displays a sharp threshold at approximately
500~\gev{}. At the threshold, production is significantly enhanced by QCD 
bound-state effects~\cite{Farrell:2006xe}. 
The result is a broad maximum that extends to 
about a \tev. At linear colliders, with an instantaneous luminosity
that grows approximately linearly with the center-of-mass energy, 
the optimum energy is typically somewhat higher than the maximum of the
cross section.

\begin{figure}[htb]
\begin{center}
\includegraphics[width=0.6\linewidth]{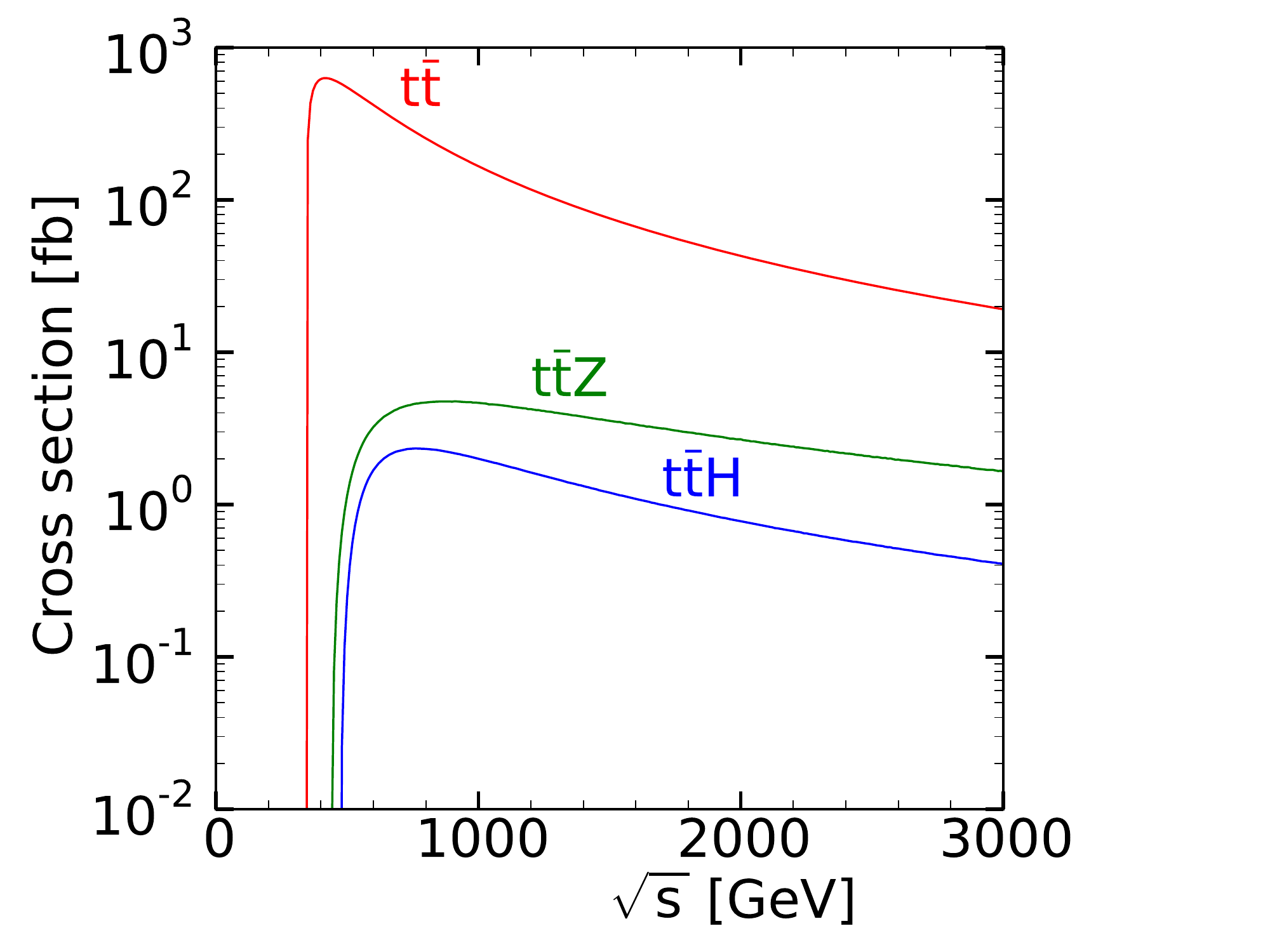}
\caption{The cross section as a function of the center-of-mass energy of lepton colliders. The three curves correspond to top quark pair production (the upper, red curve) and to associated production of a top quark pair with a $Z$-boson (the central, green curve) or a Higgs boson (the lowest, blue curve). }
\label{fig:tth_xsec}
\end{center}
\end{figure}

Philipp Roloff reviewed the ILC and CLIC studies of $t\bar{t}H$ production.
The two projects have performed full-simulation 
studies~\cite{Price:2014oca,Redford:1690648,Redford:1982243} at
several center-of-mass energies, from as low as 500~\gev{} to 1.4~\tev{}. 
Several final states are analysed. The jet multiplicity and combinatorics 
ranges from moderate (i.e. $e^+e^- \rightarrow t\bar{t}H  \rightarrow l^+ \nu b l^- \bar{\nu} \bar{b} b \bar{b}$ with four b-jets) to extremely challenging 
(i.e. $e^+e^- \rightarrow t\bar{t}H  \rightarrow q \bar{q}' b q'' \bar{q}''' WW \rightarrow 10$ jets). According to the ILC studies a precision of 18\% can 
be achieved with 500~\ifb{} at $\sqrt{s}=$ 500~\gev{}. After the complete 
programme, which includes 4000~\ifb{} at $\sqrt{s}=$ 500~\gev{}, the
precision improves to 6\%. Operation of the
ILC at $\sqrt{s} =$ 550~\gev{} (with the same luminosity) would improve
the prospects considerably: the $e^+e^- \rightarrow t\bar{t}H$ cross section
increases by a factor of nearly four and the statistical precision by
a factor two (i.e. to 9\% after 500~\ifb{} and 3\% after the nominal programme). At still higher energy there are several competing 
effects (cross section, instantaneous luminosity, experimental response).
The expected precision after 1\iab{} at 1~\tev{}~\cite{Price:2014oca}
or 1.5~\iab{} at 1.4~\tev{} in CLIC~\cite{clichiggspaper,Redford:1690648,Redford:1982243} turns out to be very similar, with an uncertainty on the Yukawa
coupling of approximately 4\%.

\begin{table}
\centering
\caption{The expected precision on the top quark Yukawa coupling extraction from the cross section for associated production of a top quark pair with a Higgs boson. For the LHC measurement at $\sqrt{s}=$ 8~\tev the value for $\mu_{ttH}$ is given, the ratio of the measured associated production rate and the Standard Model prediction.}
\begin{tabular}{l|ccc|ccc|c}
\hline
Collider & \multicolumn{3}{|c|}{LHC} & \multicolumn{3}{|c|}{ILC} & CLIC \\
$\sqrt{s}$ [TeV] & 8   & 14         & 14         & 0.5 & 0.55    & 1        & 1.4  \\        
$\int{L}$ [\ifb] & 20  & 300        & 3000       & 500 (4000) & 500 (4000) & 1000     & 1500 \\   
source           &  \cite{ATLAS-CONF-2015-044}   &  \cite{Agashe:2013hma}         &    \cite{Agashe:2013hma}        & \cite{Fujii:2015jha} & \cite{Fujii:2015jha}  &   \cite{Price:2014oca} &  \cite{clichiggspaper,Redford:1690648,Redford:1982243} \\ \hline
precision [\%]   &   $\mu =$ 2.3$^{+0.7}_{-0.6}$   &  14-15   & 7-10        &  18 (6)  & 9 (3)   & 4  & 4 \\ \hline
\end{tabular}
\label{tab:polemass}
\end{table}

The linear collider prospect studies have so far been limited to 
extractions of the Yukawa coupling from the total cross section for
$t\bar{t}H$ production. More sophisticated analyses may find better
constraints by analyzing differential distributions. These may
also provide more insight into the potential for a determination 
of other properties of the Higgs boson (for instance its CP
structure~\cite{Godbole:2011hw}).

\subsection{Yukawa coupling from a \ttbar{} threshold scan}

The top quark pair production process is sensitive to the exchange of a 
Higgs boson between the top and anti-top quark. According to Ref.~\cite{Horiguchi:2013wra}
this leads to a 9\% effect on the cross section, that is approximately
independent of the center-of-mass energy in the vicinity of the threshold. 
A precise measurement of the cross section then allows for an
extraction of the Yukawa coupling. In Ref.~\cite{Horiguchi:2013wra} the top quark
properties are extracted in a simultaneous fit of the top quark mass,
the top quark width and the Yukawa coupling to several distributions
(cross section, $A_{FB}$, top quark momentum) in the usual ten-point
threshold scan. The fit yields a statistical uncertainty on the Yukawa 
coupling of 4.2\% (assuming two different polarizations).

The authors of Ref.~\cite{Beneke:2015lwa}
scrutinize the theory uncertainty of this measurement.
A propagation of the current theory uncertainty would yield an uncertainty
on the Yukawa coupling of approximately 30\%. 
(ignoring the parametric error due
to the uncertainty in the strong coupling constant $\alpha_s$). 
Further progress in theory is therefore required to fully take advantage
of the statistical power of the threshold scan.
In the words of Ref.~\cite{Beneke:2015lwa} 
``once theoretical uncertainties are taken into account, it is unlikely 
that such a high precision [i.e. 4.2\%] can be achieved'' in practice.
That said, the extraction of the top quark Yukawa coupling from
the threshold scan remains an interesting possibility, that should
be pursued. This is particularly true for circular machines, where 
the \ttbar{} threshold may be accessible if a large enough ring
is built, while the $t\bar{t}H$ production process
seems out of reach. In that case the threshold scan provides the most direct
access to the top quark Yukawa coupling.

\subsection{Summary}
The top quark Yukawa 
coupling can be determined to 4\% statistical uncertainty, 
at the \ttbar{} production threshold. The total uncertainty 
of this measurement is expected to be approximately 20\%, 
dominated by the uncertainty in the cross section in the
threshold region.
Observation of the associated production of a top quark pair with a Higgs
boson is possible at lepton colliders with a center-of-mass energy greater
than 500~\gev{}. In the nominal luminosity scenarios ILC and CLIC can
provide a very competitive precision of approximately 4\%. This direct
extraction of the top quark Yukawa coupling thus approaches (but does not quite
reach) the precision of the constraint on the top quark Yukawa coupling 
that can be obtained indirectly in a seven-parameter
fit with $\kappa_u = \kappa_c = \kappa_t$.

\section{Top quark electro-weak couplings}
\label{sec:ttz}
The measurement of the couplings of quarks to neutral electro-weak gauge 
boson at lepton colliders has proven to be a powerful probe of new physics. 
Tight constraints on extensions of the Standard Model can be derived from 
the precision measurements of the $Zb\bar{b}$ coupling at LEP and SLD.
The study of the top quark pair production process in $e^+e^-$ collisions
finally extends this precision programme to the top quark sector.

\subsection{Impact of BSM physics}

The couplings of the top quark to the Z-boson and the photon 
are very sensitive to effects from massive unknown particles.
A precise characterization of the $Zt\bar{t}$ vertex 
is therefore a powerful handle to discover new physics 
at a scale well beyond the direct reach of the machine,
or to constrain extensions of the Standard Model.
The possible impact of several new physics scenarios is illustrated
in Fig.~\ref{fig:models}, that was shown by Stefania de Curtis in her
overview of composite Higgs models and their imprint on top physics. 
Large deviations - up to tens of \% - are allowed
in the left-handed and right-handed coupling of the top quark to the 
$Z$-boson. The large, purple markers indicate several models with extra spatial dimensions collected in Ref.~\cite{Richard:2014upa}. The cloud of smaller, black markers correspond to different realizations of the model of Ref.~\cite{Barducci:2015aoa}.

The sensitivity that can be gained through the study of 
associated production of a top quark pair with a $Z$-boson at the LHC 
is indicated by the 
large shaded ellipse, following the study of Ref.~\cite{Baur:2004uw}. The 
much tighter constraints from lepton colliders is indicated by the 
small ellipses at the origin. The estimates for the experimental studies 
from Refs.~\cite{Amjad:2015mma,Baur:2004uw} are detailed in the following 
sections.

\begin{figure}[htb]
\begin{center}
\includegraphics[width=0.8\linewidth]{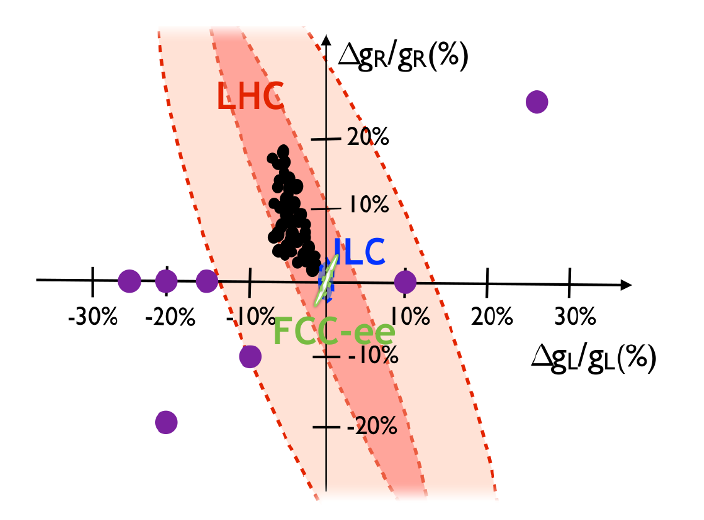}
\caption{Deviations from the SM predictions of the left-handed and right-handed couplings of the top quark to the $Z$-boson in several BSM scenarios. The large, purple markers indicate several models with extra spatial dimensions collected in Ref.~\cite{Richard:2014upa}. The cloud of smaller, black markers correspond to different realizations of the model of Ref.~\cite{Barducci:2015aoa}. The experimental prospects are discussed in the text.}
\label{fig:models}
\end{center}
\end{figure}

\subsection{LHC status and prospects}

Hadron colliders gain sensitivity to the couplings of the top quark to
neutral electro-weak gauge bosons through the study of the associated 
production processes $pp (p\bar{p}) \rightarrow t \bar{t} Z$ and 
$pp (p\bar{p}) \rightarrow  t \bar{t} \gamma$. The production cross sections
for these processes have proven to be prohibitively small at the Tevatron.
At the LHC the Standard Model predicts cross sections of order 100~fb. 
Indeed, after Run I of the LHC at 7/8~\tev{} the ATLAS and CMS experiments
have isolated the signal of both processes (and of the 
$ pp \rightarrow t \bar{t} W$ process) with a significance of greater
than 5~$\sigma$~\cite{Aad:2015eua,Khachatryan:2015sha,Aad:2015uwa,CMS-PAS-TOP-13-011}. The first measurements of the production rates are in good agreement
with the Standard Model expectations. CMS has taken advantage of 
this observation to present the first preliminary limits on the top quark 
couplings to the $Z$-boson.

Many measurements have been performed of further electro-weak processes
involving top quarks and the $W$-boson, such as single top quark production, 
and top quark decay ($t\rightarrow Wb$). The interpretation of all 
measurements in a global fit to a complete set of dimension-6 operators
related to the top sector is taking off~\cite{Buckley:2015nca,Bernreuther:2015yna,Buckley:2015lku,Rosello:2015sck,Bylund:2016phk}.
The results of Ref.~\cite{Buckley:2015lku} include (still weak) 
limits on electro-weak operators, even if they are not yet fully incorporated
into the global fit framework.

While the early observation of the associated production processes bodes well
for the remainder of the LHC programme, current constraints are quite weak.
So far, the LHC collaborations have 
refrained from presenting prospects for the sensitivity of these measurements 
in Run II and after a luminosity upgrade of the LHC. Therefore, the results
of Ref.~\cite{Baur:2004uw} remain a reference. Recent work by R\"ontsch and
Schulze~\cite{Rontsch:2014cca} shows that inclusion of the NLO correction 
for the $t\bar{t}Z$ production process reduces the uncertainties by 25$-$40\%. 
Assuming a residual theoretical uncertainty
of 15\% at NLO they estimate that with 300~\ifb{} of data at 13~\tev{} 
the vector and axial couplings can be constrained to $C_V =$ 0.24$^{+0.39}_{-0.85}$
 and $C_A =$ $-$0.6$^{+0.14}_{-0.18}$ at 95\% confidence level (the central values
are set to the SM prediction).

\subsection{Status of SM predictions}

Predictability is a key ingredient of the precision physics programme 
of lepton colliders. The rate of the 
$e^+e^- \rightarrow Z/\gamma^* \rightarrow t \bar{t}$ process 
is predicted with \%-level precision already today
(see Ref.~\cite{Amjad:2013tlv} and references therein). Calculation
of further orders in $\alpha_s$ can bring the QCD corrections to the per-mil
level. This should be compared to the uncertainty of approximately 4\% 
on the \ttbar{} cross section at the LHC~\cite{Czakon:2013goa}.

At the workshop Nhi Quach presented the status of the ongoing effort of
the GRACE collaboration to determine higher-order electro-weak
corrections to top quark pair and $t\bar{t}\gamma$
production at lepton colliders with polarized 
beams~\cite{Khiem:2012bp,Khiem:2015ofa}. The EW correction is
sizeable not only on the total cross section, but can also
affect differential results, such as the forward-backward asymmetry.
Using the narrow-width approximation the EW corrections can be
determined also for the $e^+e^- \rightarrow \ttbar \rightarrow W^+bW^-\bar{b}$
including top quark decays.

To accomodate for NLO QCD corrections,
the WHIZARD generator uses external virtual matrix element from one-loop 
providers, while providing real radiation and subtraction terms for a finite 
integration internally. Processes relevant for top physics that have 
been scrutinized already are $e^+e^- \to tt$ and $e^+e^- \to tth$ and
some the 2 $\to$ 3, 2 $\to$ 4, 2 $\to$ 5 and 2 $\to$ 6 processes that arise
when the decays are included (i.e. $tWb, WbWb, WbWbh, bb\ell\nu\ell\nu$).
In the context of this summary, the WHIZARD generator provides a 
description of the full 
$e^+e^- \rightarrow W^+bW^-\bar{b}$ process, including diagrams with a
single top quark and diagrams without $Wb$ resonance, 
at Next-to-Leading-Order in QCD~\cite{Weiss:2015npa}.
The integration of this calculation in a matrix element
generator allows for evaluation of arbitrary differential
distributions and a (POWHEG-)matching to the parton shower. The latter is 
again provided in WHIZARD for arbitrary processes.

\subsection{Linear Collider prospects}

At lepton colliders running well above the top quark pair 
production threshold the dominant top quark production process 
 is $e^+e^- \rightarrow Z/\gamma^* \rightarrow t \bar{t}$. With the
decay $t \rightarrow W b$, and $W \rightarrow q\bar{q}'$ or 
$W \rightarrow l \nu$ this is indeed the dominant six-fermion process.

The potential for the measurement of the CP conserving form factors of
the $t\bar{t} Z$ and $t\bar{t}\gamma$ vertices is well 
established by studies that include a full simulation of the
detector response in a realistic environment~\cite{Amjad:2015mma}
at the ILC operated at 500~\gev. A precise measurement of the cross section
and the forward-backward asymmetry for data taken with $e^-_Le^+_R$ 
and $e^-_Re^+_L$ polarization allows to constrain the vectorial
and axial form factors of the $Z-$boson and photon to sub-\% 
precision~\cite{Amjad:2015mma}, exceeding the expected precision
of the LHC by an order of magnitude or more. The potential remains 
excellent at energies closer to the pair production threshold.  
Only the sensitivity to the $F_{1A}$ form factor is degraded considerably
due to the reduced boost of the top quarks.
 
Roman P\"oschl presented preliminary results from the IFIC-LAL team for
CP violating couplings. These can be extracted from the asymmetries in 
observables proposed by 
Bernreuther et al.~\cite{Bernreuther:1996jk,bib:cpvbernreuther2} 
(in lepton+jets events
observables are used that are based on the directions of the charged 
lepton, the recoiling hadronic top system and the incoming electron beam).
Experimental uncertainties on these observables are expected to be
smaller than the statistical uncertainty after collecting 500~\ifb{}, thus
validating the prospects of \%-level determinations of the real and imaginary
parts of $F_{2A}$ derived at parton-level from the TESLA 
TDR~\cite{Richard:2001qm} study~\cite{AguilarSaavedra:2001rg}.
 
Alternative approaches to the measurement of top quark couplings 
are being pursued by several groups~\cite{Khiem:2015ofa,Janot:2015yza}. 
Both groups have presented promising results in parton-level 
studies, where they show that a comparison of the observed final state 
to the full matrix element can simultaneously constrain all form factors
to good precision.

\subsection{Summary}
The couplings of the top quark to the $Z$-boson and photon are a flagship
measurement of any lepton collider that is able to produce top quark pairs.
The sub-\% precision on anomalous couplings yields sensitivity to 
new physics at scales well beyond the direct reach of the machine.
The top quark pair production process moreover presents a 
sensitive probe for CP violation in the top quark sector.


\section{Exotic top quark decays} 
\label{sec:fcnc}

The potential of hadron colliders for the search for exotic flavour-changing 
neutral current decays of the top quark is well established (for a recent 
overview see Ref.~\cite{Agashe:2013hma}). The prospects of high-energy 
lepton collider experiments 
are much less explored. The rate of top quark pair production clearly 
favours the LHC. It is therefore expected that for decays where the 
final state presents distinctive signatures that are readily spotted
among the large backgrounds the LC projects are not competitive. 
They can, however, provide quite competitive limits in some cases. Lepton
colliders offer the possibility to study the FCNC coupling also in 
production (i.e. $e^+e^- \rightarrow Z/\gamma^* \rightarrow t\bar{q}$).
A second advantage is the much cleaner environment (lower background rates,
better detector performance). This may give the LC a clear advantage
if the top quark decays to final states with less
distinctive features. In the following we evaluate one example, 
the decay $t\rightarrow cH$, followed by the dominant decay 
$H \rightarrow b \bar{b}$, in some detail.

\subsection{FCNC top decays in two Higgs doublet models}

In the Standard Model FCNC top decays are strongly suppressed by the GIM
mechanism and the CKM matrix. 
Typical branching fractions for $t \rightarrow Z q $,
$ t \rightarrow g q$, $ t \rightarrow \gamma q$ and $t \rightarrow H q$ 
are of the order $10^{-15} - 10^{-12}$. In extensions of the Standard Model 
these branching fractions can be much enhanced and may be detectable 
at collider experiments. Scenarios can be found in the most popular BSM 
families (supersymmetry, models with additional spatial dimensions), 
where the branching fractions for these decays is enhanced by orders of 
magnitude. In the models considered in Ref.~\cite{Agashe:2013hma} the
decay the FCNC branching fraction is largest for $t\rightarrow cH$ 
decay for several two-Higgs-doublet models and the Randall Sundrum
model with warped extra dimensions\footnote{The exception is the R-parity violating SUSY model, where the preferred channels are $Zq$ and $gq$.} 
The authors quote a maximum branching fraction of order 0.1\% in the 
flavour-violating 2HDM model. 

Gauhar Abbas presented an investigation into the possible enhancement of the branching ratios of $ t \rightarrow cH$ decay within the aligned two Higgs doublet model (A2HDM). Assuming that the 125 GeV Higgs-like
boson corresponds to the lightest CP-even state $h$ of the CP-conserving A2HDM and taking into account constraints coming from the measurements of the 125 GeV Higgs properties, searches for a light charged Higgs via top decays, and the flavour physics, the $ t \rightarrow cH$ fraction remains well below the expected sensitivity of the LHC and ILC, across all of the parameter space considered~\cite{Abbas:2015cua}. 

Miguel Nebot presented allowed ranges of branching ratios for the decays
$ t \rightarrow Hc$ and $H \rightarrow \tau e, \tau \mu$ in a class of two 
Higgs doublet models (by Branco, Grimus, Lavoura, or BGL) 
where flavour changing neutral scalar currents occur 
at tree level~\cite{Botella:2015hoa}. In such models flavour violating 
top and Higgs decays can occur at discovery level at per mil or even percent
level, within reach of the LHC and future colliders.

\subsection{LHC status and prospects}

Current limits from the LHC (after analysis of the 8~\tev{} data set) on
flavour changing neutral current top quark decays range from a few \%
for decays to a photon and a quark to 3 $\times 10^{-5}$ for 
$ t \rightarrow gu$ (derived from a search for $qg \rightarrow t$ production). 
The ATLAS and CMS experiments have presented limits on the branching fraction
BR$(t \rightarrow cH) <$ of 0.56\% (CMS) and 0.79\% (ATLAS).

The prospects for improvements in the next two decades are presented in 
Ref.~\cite{Agashe:2013hma}. The completion of the LHC programme and its
luminosity upgrade is expected to yield improvements in the limits to
the level of $10^{-6} - 10^{-4}$. An extrapolation yields an expected limit
on BR$(t \rightarrow cH)$ of 5 $\times 10^{-4}$.

\subsection{LC prospects}

The TESLA studies reported in Ref.~\cite{AguilarSaavedra:2001ab} provide
an estimate of the expected sensitivity to $t \rightarrow \gamma q$ and
$t \rightarrow Z q$ of a linear collider experiment that collects 500~\ifb{}
at $\sqrt{s} = $ 500~\gev. The sensitivity of searches for single top 
production in association with a light quark is quite competitive with that
of the full LHC programme, reaching a BR of 6.4 $\times 10^{-6}$ for the
branching fraction to a photon and a quark. The authors of 
Ref.~\cite{Agashe:2013hma} note that a competitive limit is possible
even with the low-energy stage at 250~\gev{} of a linear collider.

\subsection{FCNC decay $t \rightarrow cH$}

The prospects for the decay $t \rightarrow cH$ are least solidly established.
The estimate for the LHC in Ref.~\cite{Agashe:2013hma} is based on an 
extrapolation, while no results are presented for the LC potential.
At the TopLC15 workshop A.F.~\.Zarnecki presented a recent LC study at 
parton level. The $t \rightarrow cH$ signal is isolated among the
large background of top quark pairs with $t \rightarrow W b$ decays in
the lepton+jets and fully hadronic final states by a series of cuts,
including the requirement of 3 b-tagged jets and a comparison of the
$\chi^2$ values of a kinematic fit to the signal and background hypotheses.
Even if this study is based on a simplified description
of the detector the author has shown that the result is relatively robust
against a degradation of the assumed jet energy resolution and flavour
tagging performance of the experiment. The expected 95\% C.L. limits
 for three different center-of-mass energies are presented as a function 
of integrated luminosity in Figure~\ref{fig:limitstch}. In this result,
an energy resolution of 50\%/$\sqrt{E [GeV]}$ is assumed.

\begin{figure}[htb]
\begin{center}
\includegraphics[width=0.8\linewidth]{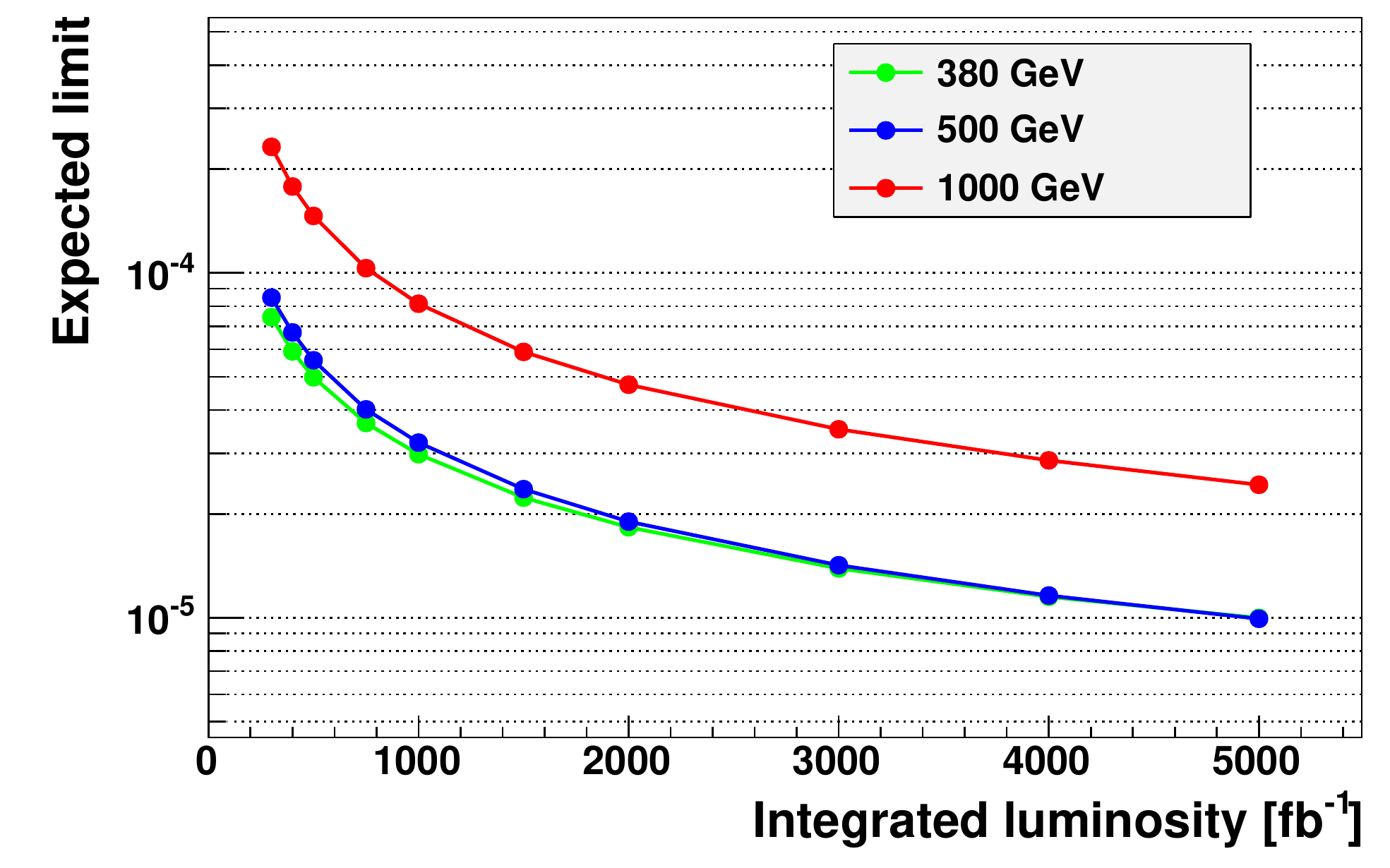}
\caption{The sensitivity of lepton colliders for the FCNC decay $t \rightarrow cH$. The expected limit on the branching fraction $t \rightarrow cH$ is shown as a function of integrated luminosity for $e^+e^-$ colliders operated at 380~\gev{}, 500~\gev{} and 1~\tev.}
\label{fig:limitstch}
\end{center}
\end{figure}

The expected limits improve roughly inversely proportional to the number of
top quarks produced in the experiment. For this reason, center-of-mass 
energies close to the maximum of the cross-section 
(i.e. approximately 420~\gev{}) offer greater potential. The larger
instantaneous luminosity that can be achieved at higher-energy
linear colliders does not make up for the loss in cross section.

\subsection{Summary}

Lepton colliders may provide competitive constraints on the rates 
of exotic decays of the top quark even after the full LHC programme. 
Preliminary, parton-level results indicate that ILC or CLIC
can achieve a sensitivity for the $t \rightarrow cH$ decay
down to a branching of $10^{-5}$.

\section{Reconstruction}
\label{sec:reco}

The top quark physics programme at lepton colliders poses stringent 
requirements on the performance of the detector and the reconstruction 
software. The ILD and SiD detector 
concepts~\cite{Abe:2010aa,Aihara:2009ad,Behnke:2013lya} for the ILC 
and the CLIC detector~\cite{Linssen:2012hp}
provide a detailed detector model in GEANT4~\cite{Agostinelli:2002hh}. 
These designs are based on decades of experience and are backed up 
by characterizations of the performance of key technologies in beam tests.

In an overview talk at the workshop Jenny List identified the key challenges 
in the development of reconstruction algorithms for top physics:
jet reconstruction, flavour tagging and 
vertex charge reconstruction. This overview was followed 
by focused contributions by Junping Tian, 
Sviatoslav Bilokin and Masakazu Kurata.  
 
Jet clustering at future lepton colliders is a considerable challenge.
Reconstruction of the six- or eight-fermion final states that arises from 
the decay of a top quark pair or $t\bar{t}H$ event requires excellent 
clustering. The higher $Q^2$ available at high-energy colliders 
increases the depth of the parton shower: the distance within a given
shower may well exceed the distance between two jets. 
At the same time, the presence of {\em pile-up} due to the
$\gamma \gamma \rightarrow$ {\em hadrons} production forces to carefully
preselect tracks and clusters (see, for instance, Refs.~\cite{Durig:2014lfa}
and~\cite{Marshall:2012ry} for, respectively, the ILC and CLIC case) 
and to limit the exposed jet area, especially in the forward region
of the experiment. 
Most analyses have resorted to the longitudinally invariant $k_t$ 
algorithm~\cite{Catani:1993hr,Ellis:1993tq} to provide more
robust clustering, but several groups have proposed new 
algorithms~\cite{Boronat:2014hva,Georgi:2014zwa,Stewart:2015waa}.
An exhaustive comparison of the performance is still lacking.

The performance of flavour tagging and vertex charge determination are
crucial to top physics. The former is key in isolating the \ttbar{} signal,
the latter provides a tag of top and anti-top quarks in fully hadronic
events. Excellent track reconstruction, with high efficiency and 
negligible fake rate, lies at the heart of both. The LC detector concepts
have demonstrated good performance in a realistic environment for prompt,
central tracks with a momentum greater than 1~\gev. The main challenge
is to extend the pattern recognition performance to low-momentum tracks
in the forward detector, that are too often missed by the pattern
recognition. A recovery procedure presented by Sviatoslav Bilokin
has shown some potential to improve the vertex charge determination,
but more work is needed to improve the track reconstruction for this
category of particles.

\section{Summary and outlook}
\label{sec:summary}

The TopLC15 workshop at IFIC in Valencia in July 2015 brought together 
the theory and experimental communities with an interest in the top physics 
of future high-energy lepton colliders. 
This summary of the contributions provides an essentially complete status
report from ongoing studies into the potential of the linear $e^+e^-$ collider
projects (ILC and CLIC) to measure top quark properties and to study the
interactions of the top quark with other Standard Model particles.

The case for a top quark mass measurement at threshold 
with a statistical precision of
order 20~\mev{} and a total uncertainty of less than 50~\mev rests on 
increasingly solid ground. This precision includes the theory uncertainty 
- based on today's state-of-the-art NNNLO description of top quark
pair production at threshold and the four-loop conversion of the 
threshold mass to the $\bar{MS}$ mass. Also experimental uncertainties
are estimated. The workshop saw a renewed interest in mass measurements
in the continuum, where the mass may be extracted from the cross section
for associated $\ttbar + $ photon or $\ttbar + $ jet production.

The top quark Yukawa coupling can be determined at threshold, to a precision
that is likely limited by theory uncertainties ($\sim$20\% with the 
current theory). Colliders operated above the $t \bar{t} H$ threshold 
at approximately 500-550~\gev can achieve a precision of approximately 4\%.

The measurement of the top quark electro-weak couplings are another 
pillar of the top physics programme. Sizeable (O(10\%)) deviations from the
Standard Model predictions are expected in a broad range of extensions 
of the SM, whereas a percent-level measurement is feasible with the nominal
ILC and CLIC programmes. 

Lepton colliders can provide competitive limits on rare top quark decays
that are not easily distinguished from the dominant (hadronic) backgrounds
at the LHC. A good example is the preliminary result of a parton-level study 
that predicts a sensitivity to $ t \rightarrow c H$ to a branching fraction
of the order of $10^{-5}$.

\bigskip
\section*{Acknowledgements}
We acknowledge the financial support of the regional government - the Generalitat Valenciana - for TopLC15 (AORG/2015/113) and the support of IFIC under the Severo Ochoa programme (SEV-2014-0398).


\bibliographystyle{JHEP}
\bibliography{confnotebib_atlas}

\end{document}

%% file: econfmacros.tex
\def\beq{\begin{equation}}
\def\eeq#1{\label{#1}\end{equation}}
\def\eeqn{\end{equation}}

\def\beqa{\begin{eqnarray}}
\def\eeqa#1{\label{#1}\end{eqnarray}}
\def\eeqan{\end{eqnarray}}

\let\bar=\overbar

\def\Dslash{\not{\hbox{\kern-4pt $D$}}}
\def\dslash{\not{\hbox{\kern-2pt $\del$}}}

\def\msb{{\bar{\ssstyle M \kern -1pt S}}}




%% file: article.bbl
\providecommand{\href}[2]{#2}\begingroup\raggedright\begin{thebibliography}{10}

\bibitem{Baer:2013cma}
H.~Baer, T.~Barklow, K.~Fujii, Y.~Gao, A.~Hoang, et~al., {\it {The
  International Linear Collider Technical Design Report - Volume 2: Physics}},
  \href{http://arxiv.org/abs/1306.6352}{{\tt arXiv:1306.6352}}.

\bibitem{Linssen:2012hp}
L.~Linssen, A.~Miyamoto, M.~Stanitzki, and H.~Weerts, {\it {Physics and
  Detectors at CLIC: CLIC Conceptual Design Report}},
  \href{http://arxiv.org/abs/1202.5940}{{\tt arXiv:1202.5940}}.

\bibitem{Gomez-Ceballos:2013zzn}
M.~Bicer, H.~Duran~Yildiz, I.~Yildiz, G.~Coignet, M.~Delmastro, et~al., {\it
  {First Look at the Physics Case of TLEP}},
  \href{http://arxiv.org/abs/1308.6176}{{\tt arXiv:1308.6176}}.

\bibitem{CEPC-SPPCStudyGroup:2015csa}
{CEPC-SPPC Study Group}, {\it {CEPC-SPPC Preliminary Conceptual Design Report.
  1. Physics and Detector, IHEP-CEPC-DR-2015-01 (2015)}}.

\bibitem{Alexahin:2013ojp}
Y.~Alexahin et~al., {\it {Muon Collider Higgs Factory for Snowmass 2013}},  in
  {\em {Community Summer Study 2013: Snowmass on the Mississippi (CSS2013)
  Minneapolis, MN, USA, July 29-August 6, 2013}}, 2013.
\newblock \href{http://arxiv.org/abs/1308.2143}{{\tt arXiv:1308.2143}}.

\bibitem{Dawson:2013bba}
S.~Dawson et~al., {\it {Working Group Report: Higgs Boson}},  in {\em
  {Community Summer Study 2013: Snowmass on the Mississippi (CSS2013)
  Minneapolis, MN, USA, July 29-August 6, 2013}}, 2013.
\newblock \href{http://arxiv.org/abs/1310.8361}{{\tt arXiv:1310.8361}}.

\bibitem{Gershtein:2013iqa}
Y.~Gershtein et~al., {\it {Working Group Report: New Particles, Forces, and
  Dimensions}},  in {\em {Community Summer Study 2013: Snowmass on the
  Mississippi (CSS2013) Minneapolis, MN, USA, July 29-August 6, 2013}}, 2013.
\newblock \href{http://arxiv.org/abs/1311.0299}{{\tt arXiv:1311.0299}}.

\bibitem{Agashe:2013hma}
{\bf Top Quark Working Group} Collaboration, K.~Agashe et~al., {\it {Working
  Group Report: Top Quark}},  in {\em {Community Summer Study 2013: Snowmass on
  the Mississippi (CSS2013) Minneapolis, MN, USA, July 29-August 6, 2013}},
  2013.
\newblock \href{http://arxiv.org/abs/1311.2028}{{\tt arXiv:1311.2028}}.

\bibitem{Price:2014oca}
T.~Price, P.~Roloff, J.~Strube, and T.~Tanabe, {\it {Full simulation study of
  the top Yukawa coupling at the ILC at $\sqrt{s}=$ 1 TeV}},  {\em Eur. Phys.
  J.} {\bf C75} (2015), no.~7 309, [\href{http://arxiv.org/abs/1409.7157}{{\tt
  arXiv:1409.7157}}].

\bibitem{Amjad:2015mma}
M.~S. Amjad et~al., {\it {A precise characterisation of the top quark
  electro-weak vertices at the ILC}},  {\em Eur. Phys. J.} {\bf C75} (2015),
  no.~10 512, [\href{http://arxiv.org/abs/1505.06020}{{\tt arXiv:1505.06020}}].

\bibitem{Baak:2014ora}
{\bf Gfitter Group} Collaboration, M.~Baak, J.~Cúth, J.~Haller, A.~Hoecker,
  R.~Kogler, K.~Mönig, M.~Schott, and J.~Stelzer, {\it {The global electroweak
  fit at NNLO and prospects for the LHC and ILC}},  {\em Eur. Phys. J.} {\bf
  C74} (2014) 3046, [\href{http://arxiv.org/abs/1407.3792}{{\tt
  arXiv:1407.3792}}].

\bibitem{Degrassi:2012ry}
G.~Degrassi, S.~Di~Vita, J.~Elias-Miro, J.~R. Espinosa, G.~F. Giudice,
  G.~Isidori, and A.~Strumia, {\it {Higgs mass and vacuum stability in the
  Standard Model at NNLO}},  {\em JHEP} {\bf 08} (2012) 098,
  [\href{http://arxiv.org/abs/1205.6497}{{\tt arXiv:1205.6497}}].

\bibitem{ATLAS:2014wva}
{\bf ATLAS, CDF, CMS, D0} Collaboration, {\it {First combination of Tevatron
  and LHC measurements of the top-quark mass}},
  \href{http://arxiv.org/abs/1403.4427}{{\tt arXiv:1403.4427}}.

\bibitem{Beneke:2000hk}
M.~Beneke et~al., {\it {Top quark physics}},  in {\em {1999 CERN Workshop on
  standard model physics (and more) at the LHC, CERN, Geneva, Switzerland,
  25-26 May: Proceedings}}, 2000.
\newblock \href{http://arxiv.org/abs/hep-ph/0003033}{{\tt hep-ph/0003033}}.

\bibitem{Juste:2013dsa}
A.~Juste, S.~Mantry, A.~Mitov, A.~Penin, P.~Skands, E.~Varnes, M.~Vos, and
  S.~Wimpenny, {\it {Determination of the top quark mass circa 2013: methods,
  subtleties, perspectives}},  {\em Eur. Phys. J.} {\bf C74} (2014), no.~10
  3119, [\href{http://arxiv.org/abs/1310.0799}{{\tt arXiv:1310.0799}}].

\bibitem{CMS-PAS-FTR-13-017}
{\bf CMS} Collaboration, {\it {Projected improvement of the accuracy of
  top-quark mass measurements at the upgraded LHC}},  {\em CMS-PAS-FTR-13-017}.

\bibitem{CMS-PAS-TOP-13-004}
{\bf CMS} Collaboration, {\it {Measurement of the \ttbar{} production cross
  section in the $e\mu$ channel in pp collisions at 7 and 8 TeV}},  {\em
  CMS-PAS-TOP-13-004}.

\bibitem{Aad:2014kva}
{\bf ATLAS} Collaboration, {\it {Measurement of the $t\overline{t}$ production
  cross-section using $e\mu $ events with $b$ -tagged jets in $pp$ collisions
  at $\sqrt{s}=7$ and 8 TeV with the ATLAS detector}},  {\em Eur. Phys. J.}
  {\bf C74} (2014), no.~10 3109, [\href{http://arxiv.org/abs/1406.5375}{{\tt
  arXiv:1406.5375}}].

\bibitem{Czakon:2013goa}
M.~Czakon, P.~Fiedler, and A.~Mitov, {\it {Total Top-Quark Pair-Production
  Cross Section at Hadron Colliders Through $O(α\frac{4}{S})$}},  {\em Phys.
  Rev. Lett.} {\bf 110} (2013) 252004,
  [\href{http://arxiv.org/abs/1303.6254}{{\tt arXiv:1303.6254}}].

\bibitem{Aad:2015waa}
{\bf ATLAS} Collaboration, {\it {Determination of the top-quark pole mass using
  $t \bar{t}+1$-jet events collected with the ATLAS experiment in 7 TeV $pp$
  collisions}},  \href{http://arxiv.org/abs/1507.01769}{{\tt
  arXiv:1507.01769}}.

\bibitem{Alioli:2013mxa}
S.~Alioli, P.~Fernandez, J.~Fuster, A.~Irles, S.-O. Moch, P.~Uwer, and M.~Vos,
  {\it {A new observable to measure the top-quark mass at hadron colliders}},
  {\em Eur. Phys. J.} {\bf C73} (2013) 2438,
  [\href{http://arxiv.org/abs/1303.6415}{{\tt arXiv:1303.6415}}].

\bibitem{ahoang08}
A.~H. Hoang and I.~W. Stewart, {\it {Top-mass measurements from jets and the
  Tevatron top mass}},  {\em Nouvo Cimento} {\bf B123} (2008) 1092--1100.

\bibitem{Moch:2014tta}
S.~Moch et~al., {\it {High precision fundamental constants at the TeV scale}},
  \href{http://arxiv.org/abs/1405.4781}{{\tt arXiv:1405.4781}}.

\bibitem{ahoang14}
A.~H. Hoang, {\it {The Top Mass: Interpretation and Theoretical
  Uncertainties}},  \href{http://arxiv.org/abs/1412.3649}{{\tt
  arXiv:1412.3649}}.

\bibitem{Corcella:2015kth}
G.~Corcella, {\it {Interpretation of the top-quark mass measurements: a theory
  overview}},  in {\em {8th International Workshop on Top Quark Physics
  (TOP2015) Ischia, NA, Italy, September 14-18, 2015}}, 2015.
\newblock \href{http://arxiv.org/abs/1511.08429}{{\tt arXiv:1511.08429}}.

\bibitem{Sjostrand:2006za}
T.~Sjostrand, S.~Mrenna, and P.~Z. Skands, {\it {PYTHIA 6.4 Physics and
  Manual}},  {\em JHEP} {\bf 05} (2006) 026,
  [\href{http://arxiv.org/abs/hep-ph/0603175}{{\tt hep-ph/0603175}}].

\bibitem{Gritschacher:2013pha}
S.~Gritschacher, A.~H. Hoang, I.~Jemos, and P.~Pietrulewicz, {\it {Secondary
  Heavy Quark Production in Jets through Mass Modes}},  {\em Phys. Rev.} {\bf
  D88} (2013) 034021, [\href{http://arxiv.org/abs/1302.4743}{{\tt
  arXiv:1302.4743}}].

\bibitem{Gritschacher:2013tza}
S.~Gritschacher, A.~Hoang, I.~Jemos, and P.~Pietrulewicz, {\it {Two loop soft
  function for secondary massive quarks}},  {\em Phys. Rev.} {\bf D89} (2014),
  no.~1 014035, [\href{http://arxiv.org/abs/1309.6251}{{\tt arXiv:1309.6251}}].

\bibitem{Pietrulewicz:2014qza}
P.~Pietrulewicz, S.~Gritschacher, A.~H. Hoang, I.~Jemos, and V.~Mateu, {\it
  {Variable Flavor Number Scheme for Final State Jets in Thrust}},  {\em Phys.
  Rev.} {\bf D90} (2014), no.~11 114001,
  [\href{http://arxiv.org/abs/1405.4860}{{\tt arXiv:1405.4860}}].

\bibitem{Gusken:1985nf}
S.~Gusken, J.~H. Kuhn, and P.~M. Zerwas, {\it {Threshold Behavior of Top
  Production in $e^+ e^-$ Annihilation}},  {\em Phys. Lett.} {\bf B155} (1985)
  185.

\bibitem{Beneke:2015kwa}
M.~Beneke, Y.~Kiyo, P.~Marquard, A.~Penin, J.~Piclum, and M.~Steinhauser, {\it
  {Next-to-Next-to-Next-to-Leading Order QCD Prediction for the Top Antitop
  $S$-Wave Pair Production Cross Section Near Threshold in $e^+e^-$
  Annihilation}},  {\em Phys. Rev. Lett.} {\bf 115} (2015), no.~19 192001,
  [\href{http://arxiv.org/abs/1506.06864}{{\tt arXiv:1506.06864}}].

\bibitem{Beneke:2015lwa}
M.~Beneke, A.~Maier, J.~Piclum, and T.~Rauh, {\it {Higgs effects in top
  anti-top production near threshold in $e^+e^−$ annihilation}},  {\em Nucl.
  Phys.} {\bf B899} (2015) 180--193,
  [\href{http://arxiv.org/abs/1506.06865}{{\tt arXiv:1506.06865}}].

\bibitem{Beneke:2010mp}
M.~Beneke, B.~Jantzen, and P.~Ruiz-Femenia, {\it {Electroweak non-resonant NLO
  corrections to e+ e- -> W+ W- b bbar in the t tbar resonance region}},  {\em
  Nucl. Phys.} {\bf B840} (2010) 186--213,
  [\href{http://arxiv.org/abs/1004.2188}{{\tt arXiv:1004.2188}}].

\bibitem{Hoang:2013uda}
A.~H. Hoang and M.~Stahlhofen, {\it {The Top-Antitop Threshold at the ILC: NNLL
  QCD Uncertainties}},  {\em JHEP} {\bf 05} (2014) 121,
  [\href{http://arxiv.org/abs/1309.6323}{{\tt arXiv:1309.6323}}].

\bibitem{Kilian:2007gr}
W.~Kilian, T.~Ohl, and J.~Reuter, {\it {WHIZARD: Simulating Multi-Particle
  Processes at LHC and ILC}},  {\em Eur. Phys. J.} {\bf C71} (2011) 1742,
  [\href{http://arxiv.org/abs/0708.4233}{{\tt arXiv:0708.4233}}].

\bibitem{Simon:2016htt}
F.~Simon, {\it {A First Look at the Impact of NNNLO Theory Uncertainties on Top
  Mass Measurements at the ILC}},  in {\em {International Workshop on Future
  Linear Colliders (LCWS15) Whistler, B.C., Canada, November 2-6, 2015}}, 2016.
\newblock \href{http://arxiv.org/abs/1603.04764}{{\tt arXiv:1603.04764}}.

\bibitem{Seidel:2013sqa}
K.~Seidel, F.~Simon, M.~Tesar, and S.~Poss, {\it {Top quark mass measurements
  at and above threshold at CLIC}},  {\em Eur. Phys. J.} {\bf C73} (2013),
  no.~8 2530, [\href{http://arxiv.org/abs/1303.3758}{{\tt arXiv:1303.3758}}].

\bibitem{Poss:2013oea}
S.~Poss and A.~Sailer, {\it {Luminosity Spectrum Reconstruction at Linear
  Colliders}},  {\em Eur. Phys. J.} {\bf C74} (2014), no.~4 2833,
  [\href{http://arxiv.org/abs/1309.0372}{{\tt arXiv:1309.0372}}].

\bibitem{Simon:2014hna}
F.~Simon, {\it {Perspectives for Top Quark Physics at the (I)LC}},  in {\em
  {7th International Workshop on Top Quark Physics (TOP2014) Cannes, France,
  September 28-October 3, 2014}}, 2014.
\newblock \href{http://arxiv.org/abs/1411.7517}{{\tt arXiv:1411.7517}}.

\bibitem{Fuster:2015jva}
J.~Fuster, I.~García, P.~Gomis, M.~Perelló, E.~Ros, and M.~Vos, {\it {Study
  of single top production at high energy electron positron colliders}},  {\em
  Eur. Phys. J.} {\bf C75} (2015) 223,
  [\href{http://arxiv.org/abs/1411.2355}{{\tt arXiv:1411.2355}}].

\bibitem{Marquard:2015qpa}
P.~Marquard, A.~V. Smirnov, V.~A. Smirnov, and M.~Steinhauser, {\it {Quark Mass
  Relations to Four-Loop Order in Perturbative QCD}},  {\em Phys. Rev. Lett.}
  {\bf 114} (2015), no.~14 142002, [\href{http://arxiv.org/abs/1502.01030}{{\tt
  arXiv:1502.01030}}].

\bibitem{Agashe:2014kda}
{\bf Particle Data Group} Collaboration, K.~A. Olive et~al., {\it {Review of
  Particle Physics}},  {\em Chin. Phys.} {\bf C38} (2014) 090001.

\bibitem{d'Enterria:2015toz}
D.~d'Enterria and P.~Z. Skands, eds., {\em {High-precision $\alpha_s$
  measurements from LHC to FCC-ee}}, 2015.

\bibitem{Aoki:2013ldr}
S.~Aoki et~al., {\it {Review of lattice results concerning low-energy particle
  physics}},  {\em Eur. Phys. J.} {\bf C74} (2014) 2890,
  [\href{http://arxiv.org/abs/1310.8555}{{\tt arXiv:1310.8555}}].

\bibitem{Kiyo:2015ooa}
Y.~Kiyo, G.~Mishima, and Y.~Sumino, {\it {Strong IR Cancellation in Heavy
  Quarkonium and Precise Top Mass Determination}},  {\em JHEP} {\bf 11} (2015)
  084, [\href{http://arxiv.org/abs/1506.06542}{{\tt arXiv:1506.06542}}].

\bibitem{Fleming:2007qr}
S.~Fleming, A.~H. Hoang, S.~Mantry, and I.~W. Stewart, {\it {Jets from massive
  unstable particles: Top-mass determination}},  {\em Phys. Rev.} {\bf D77}
  (2008) 074010, [\href{http://arxiv.org/abs/hep-ph/0703207}{{\tt
  hep-ph/0703207}}].

\bibitem{Fleming:2007xt}
S.~Fleming, A.~H. Hoang, S.~Mantry, and I.~W. Stewart, {\it {Top Jets in the
  Peak Region: Factorization Analysis with NLL Resummation}},  {\em Phys. Rev.}
  {\bf D77} (2008) 114003, [\href{http://arxiv.org/abs/0711.2079}{{\tt
  arXiv:0711.2079}}].

\bibitem{ATLAS-CONF-2015-044}
T.~ATLAS and C.~collaborations, {\it {Measurements of the Higgs boson
  production and decay rates and constraints on its couplings from a combined
  ATLAS and CMS analysis of the LHC pp collision data at $\sqrt{s} =$ 7 and 8
  TeV}}, .

\bibitem{Farrell:2006xe}
C.~Farrell and A.~H. Hoang, {\it {Next-to-leading-logarithmic QCD corrections
  to the cross- section sigma(e+ e- ---> t anti-t H) at 500-GeV}},  {\em Phys.
  Rev.} {\bf D74} (2006) 014008,
  [\href{http://arxiv.org/abs/hep-ph/0604166}{{\tt hep-ph/0604166}}].

\bibitem{Redford:1690648}
S.~Redford, P.~Roloff, and M.~Vogel, {\it {Physics potential of the top Yukawa
  coupling measurement at a 1.4 TeV Compact Linear Collider using the CLIC SiD
  detector}}, CLICdp-Note-2014-001.

\bibitem{Redford:1982243}
S.~Redford, P.~Roloff, and M.~Vogel, {\it {Study of the effect of additional
  background channels on the top Yukawa coupling measurement at a 1.4 TeV
  CLIC}}, CLICdp-Note-2015-001.

\bibitem{clichiggspaper}
P.~e. Roloff, {\it {Higgs Physics at the CLIC Electron-Positron Linear
  Collider}}, in preparation.

\bibitem{Fujii:2015jha}
K.~Fujii et~al., {\it {Physics Case for the International Linear Collider}},
  \href{http://arxiv.org/abs/1506.05992}{{\tt arXiv:1506.05992}}.

\bibitem{Godbole:2011hw}
R.~M. Godbole, C.~Hangst, M.~Muhlleitner, S.~D. Rindani, and P.~Sharma, {\it
  {Model-independent analysis of Higgs spin and CP properties in the process
  $e^+ e^- \to t \bar t \Phi$}},  {\em Eur. Phys. J.} {\bf C71} (2011) 1681,
  [\href{http://arxiv.org/abs/1103.5404}{{\tt arXiv:1103.5404}}].

\bibitem{Horiguchi:2013wra}
T.~Horiguchi, A.~Ishikawa, T.~Suehara, K.~Fujii, Y.~Sumino, Y.~Kiyo, and
  H.~Yamamoto, {\it {Study of top quark pair production near threshold at the
  ILC}},  \href{http://arxiv.org/abs/1310.0563}{{\tt arXiv:1310.0563}}.

\bibitem{Richard:2014upa}
F.~Richard, {\it {Present and future constraints on top EW couplings}},
  \href{http://arxiv.org/abs/1403.2893}{{\tt arXiv:1403.2893}}.

\bibitem{Barducci:2015aoa}
D.~Barducci, S.~De~Curtis, S.~Moretti, and G.~M. Pruna, {\it {Top pair
  production at a future $e^+e^-$ machine in a composite Higgs scenario}},
  {\em JHEP} {\bf 08} (2015) 127, [\href{http://arxiv.org/abs/1504.05407}{{\tt
  arXiv:1504.05407}}].

\bibitem{Baur:2004uw}
U.~Baur, A.~Juste, L.~H. Orr, and D.~Rainwater, {\it {Probing electroweak top
  quark couplings at hadron colliders}},  {\em Phys. Rev.} {\bf D71} (2005)
  054013, [\href{http://arxiv.org/abs/hep-ph/0412021}{{\tt hep-ph/0412021}}].

\bibitem{Aad:2015eua}
{\bf ATLAS} Collaboration, {\it {Measurement of the $
  t\overline{t}W $ and $ t\overline{t}Z $ production cross sections in pp
  collisions at $ \sqrt{s}=8 $ TeV with the ATLAS detector}},  {\em JHEP} {\bf
  11} (2015) 172, [\href{http://arxiv.org/abs/1509.05276}{{\tt
  arXiv:1509.05276}}].

\bibitem{Khachatryan:2015sha}
{\bf CMS} Collaboration, {\it {Observation of top quark
  pairs produced in association with a vector boson in pp collisions at $
  \sqrt{s}=8 $ TeV}},  {\em JHEP} {\bf 01} (2016) 096,
  [\href{http://arxiv.org/abs/1510.01131}{{\tt arXiv:1510.01131}}].

\bibitem{Aad:2015uwa}
{\bf ATLAS} Collaboration, {\it {Observation of top-quark pair
  production in association with a photon and measurement of the
  $t\bar{t}\gamma$ production cross section in pp collisions at $\sqrt{s}=7$
  TeV using the ATLAS detector}},  {\em Phys. Rev.} {\bf D91} (2015), no.~7
  072007, [\href{http://arxiv.org/abs/1502.00586}{{\tt arXiv:1502.00586}}].

\bibitem{CMS-PAS-TOP-13-011}
{\bf CMS Collaboration} Collaboration, {\it {Measurement of the inclusive
  top-quark pair + photon production cross section in the muon + jets channel
  in pp collisions at 8 TeV}},  Tech. Rep. CMS-PAS-TOP-13-011, CERN, Geneva,
  2014.

\bibitem{Buckley:2015nca}
A.~Buckley, C.~Englert, J.~Ferrando, D.~J. Miller, L.~Moore, M.~Russell, and
  C.~D. White, {\it {Global fit of top quark effective theory to data}},  {\em
  Phys. Rev.} {\bf D92} (2015), no.~9 091501,
  [\href{http://arxiv.org/abs/1506.08845}{{\tt arXiv:1506.08845}}].

\bibitem{Bernreuther:2015yna}
W.~Bernreuther, D.~Heisler, and Z.-G. Si, {\it {A set of top quark spin
  correlation and polarization observables for the LHC: Standard Model
  predictions and new physics contributions}},  {\em JHEP} {\bf 12} (2015) 026,
  [\href{http://arxiv.org/abs/1508.05271}{{\tt arXiv:1508.05271}}].

\bibitem{Buckley:2015lku}
A.~Buckley, C.~Englert, J.~Ferrando, D.~J. Miller, L.~Moore, M.~Russell, and
  C.~D. White, {\it {Constraining top quark effective theory in the LHC Run II
  era}},  \href{http://arxiv.org/abs/1512.03360}{{\tt arXiv:1512.03360}}.

\bibitem{Rosello:2015sck}
M.~P. Rosello and M.~Vos, {\it {Constraints on four-fermion interactions from
  the $t\bar{t}$ charge asymmetry at hadron colliders}},
  \href{http://arxiv.org/abs/1512.07542}{{\tt arXiv:1512.07542}}.

\bibitem{Bylund:2016phk}
O.~B. Bylund, F.~Maltoni, I.~Tsinikos, E.~Vryonidou, and C.~Zhang, {\it
  {Probing top quark neutral couplings in the Standard Model Effective Field
  Theory at NLO QCD}},  \href{http://arxiv.org/abs/1601.08193}{{\tt
  arXiv:1601.08193}}.

\bibitem{Rontsch:2014cca}
R.~R{\"o}ntsch and M.~Schulze, {\it {Constraining couplings of top quarks to
  the Z boson in $ t\overline{t} $ + Z production at the LHC}},  {\em JHEP}
  {\bf 07} (2014) 091, [\href{http://arxiv.org/abs/1404.1005}{{\tt
  arXiv:1404.1005}}]. [Erratum: JHEP09,132(2015)].

\bibitem{Amjad:2013tlv}
M.~S. Amjad, M.~Boronat, T.~Frisson, I.~Garcia, R.~Poschl, E.~Ros, F.~Richard,
  J.~Rouene, P.~R. Femenia, and M.~Vos, {\it {A precise determination of top
  quark electro-weak couplings at the ILC operating at $\sqrt{s}=500$ GeV}},
  \href{http://arxiv.org/abs/1307.8102}{{\tt arXiv:1307.8102}}.

\bibitem{Khiem:2012bp}
P.~H. Khiem, J.~Fujimoto, T.~Ishikawa, T.~Kaneko, K.~Kato, Y.~Kurihara,
  Y.~Shimizu, T.~Ueda, J.~A.~M. Vermaseren, and Y.~Yasui, {\it {Full
  $\mathcal{O}(\alpha)$ electroweak radiative corrections to $e^+e^-
  \rightarrow t \bar{t} \gamma$ with GRACE-Loop}},  {\em Eur. Phys. J.} {\bf
  C73} (2013), no.~4 2400, [\href{http://arxiv.org/abs/1211.1112}{{\tt
  arXiv:1211.1112}}].

\bibitem{Khiem:2015ofa}
P.~H. Khiem, E.~Kou, Y.~Kurihara, and F.~L. Diberder, {\it {Probing New Physics
  using top quark polarization in the $e^+e^- \rightarrow t \bar{t}$ process at
  future Linear Colliders}},  2015.
\newblock \href{http://arxiv.org/abs/1503.04247}{{\tt arXiv:1503.04247}}.

\bibitem{Weiss:2015npa}
C.~Weiss, B.~C. Nejad, W.~Kilian, and J.~Reuter, {\it {Automated NLO QCD
  Corrections with WHIZARD}},  {\em PoS} {\bf EPS-HEP2015} (2015) 466,
  [\href{http://arxiv.org/abs/1510.02666}{{\tt arXiv:1510.02666}}].

\bibitem{Bernreuther:1996jk}
W.~Bernreuther, A.~Brandenburg, and P.~Overmann, {\it {CP nonconservation in
  top quark production by (un)polarized e+ e- and gamma gamma collisions}},  in
  {\em {e+ e- collisions at TeV energies: The physics potential. Proceedings,
  Workshop, Annecy, France, February 4, 1995, Gran Sasso, Assergi, Italy, June
  2-3, 1995, Hamburg, Germany, August 30-September 1, 1995}}, 1996.
\newblock \href{http://arxiv.org/abs/hep-ph/9602273}{{\tt hep-ph/9602273}}.

\bibitem{bib:cpvbernreuther2}
W.~Bernreuther and P.~Overmann, {\it {Probing Higgs boson and supersymmetry
  induced CP violation in top quark production by (un)polarized electron -
  positron collisions}},  {\em Z. Phys.} {\bf C72} (1996) 461--467,
  [\href{http://arxiv.org/abs/hep-ph/9511256}{{\tt hep-ph/9511256}}].

\bibitem{Richard:2001qm}
F.~Richard, J.~R. Schneider, D.~Trines, and A.~Wagner, {\it {TESLA, The
  Superconducting Electron Positron Linear Collider with an Integrated X-ray
  Laser Laboratory, Technical Design Report Part 1}},
  \href{http://arxiv.org/abs/hep-ph/0106314}{{\tt hep-ph/0106314}}.

\bibitem{AguilarSaavedra:2001rg}
{\bf ECFA/DESY LC Physics Working Group} Collaboration, J.~A. Aguilar-Saavedra
  et~al., {\it {TESLA: The Superconducting electron positron linear collider
  with an integrated x-ray laser laboratory. Technical design report. Part 3.
  Physics at an e+ e- linear collider}},
  \href{http://arxiv.org/abs/hep-ph/0106315}{{\tt hep-ph/0106315}}.

\bibitem{Janot:2015yza}
P.~Janot, {\it {Top-quark electroweak couplings at the FCC-ee}},  {\em JHEP}
  {\bf 04} (2015) 182, [\href{http://arxiv.org/abs/1503.01325}{{\tt
  arXiv:1503.01325}}].

\bibitem{Abbas:2015cua}
G.~Abbas, A.~Celis, X.-Q. Li, J.~Lu, and A.~Pich, {\it {Flavour-changing top
  decays in the aligned two-Higgs-doublet model}},  {\em JHEP} {\bf 06} (2015)
  005, [\href{http://arxiv.org/abs/1503.06423}{{\tt arXiv:1503.06423}}].

\bibitem{Botella:2015hoa}
F.~J. Botella, G.~C. Branco, M.~Nebot, and M.~N. Rebelo, {\it {Flavour Changing
  Higgs Couplings in a Class of Two Higgs Doublet Models}},
  \href{http://arxiv.org/abs/1508.05101}{{\tt arXiv:1508.05101}}.

\bibitem{AguilarSaavedra:2001ab}
J.~A. Aguilar-Saavedra and T.~Riemann, {\it {Probing top flavor changing
  neutral couplings at TESLA}},  in {\em {5th Workshop of the 2nd ECFA - DESY
  Study on Physics and Detectors for a Linear Electron - Positron Collider
  Obernai, France, October 16-19, 1999}}, 2001.
\newblock \href{http://arxiv.org/abs/hep-ph/0102197}{{\tt hep-ph/0102197}}.

\bibitem{Abe:2010aa}
{\bf Linear Collider ILD Concept Group -} Collaboration, T.~Abe et~al., {\it
  {The International Large Detector: Letter of Intent}},
  \href{http://arxiv.org/abs/1006.3396}{{\tt arXiv:1006.3396}}.

\bibitem{Aihara:2009ad}
H.~Aihara, P.~Burrows, M.~Oreglia, E.~L. Berger, V.~Guarino, J.~Repond,
  H.~Weerts, L.~Xia, J.~Zhang, Q.~Zhang, et~al., {\it {SiD Letter of Intent}},
  \href{http://arxiv.org/abs/0911.0006}{{\tt arXiv:0911.0006}}.

\bibitem{Behnke:2013lya}
H.~Abramowicz et~al., {\it {The International Linear Collider Technical Design
  Report - Volume 4: Detectors}},  \href{http://arxiv.org/abs/1306.6329}{{\tt
  arXiv:1306.6329}}.

\bibitem{Agostinelli:2002hh}
{\bf GEANT4} Collaboration, S.~Agostinelli et~al., {\it {GEANT4: A Simulation
  toolkit}},  {\em Nucl.Instrum.Meth.} {\bf A506} (2003) 250--303.

\bibitem{Durig:2014lfa}
C.~Dürig, K.~Fujii, J.~List, and J.~Tian, {\it {Model Independent
  Determination of $HWW$ coupling and Higgs total width at ILC}},  in {\em
  {International Workshop on Future Linear Colliders (LCWS13) Tokyo, Japan,
  November 11-15, 2013}}, 2014.
\newblock \href{http://arxiv.org/abs/1403.7734}{{\tt arXiv:1403.7734}}.

\bibitem{Marshall:2012ry}
J.~S. Marshall, A.~Münnich, and M.~A. Thomson, {\it {Performance of Particle
  Flow Calorimetry at CLIC}},  {\em Nucl. Instrum. Meth.} {\bf A700} (2013)
  153--162, [\href{http://arxiv.org/abs/1209.4039}{{\tt arXiv:1209.4039}}].

\bibitem{Catani:1993hr}
S.~Catani, Y.~L. Dokshitzer, M.~Seymour, and B.~Webber, {\it {Longitudinally
  invariant $K_t$ clustering algorithms for hadron hadron collisions}},  {\em
  Nucl.Phys.} {\bf B406} (1993) 187--224.

\bibitem{Ellis:1993tq}
S.~D. Ellis and D.~E. Soper, {\it {Successive combination jet algorithm for
  hadron collisions}},  {\em Phys.Rev.} {\bf D48} (1993) 3160--3166,
  [\href{http://arxiv.org/abs/hep-ph/9305266}{{\tt hep-ph/9305266}}].

\bibitem{Boronat:2014hva}
M.~Boronat, J.~Fuster, I.~Garcia, E.~Ros, and M.~Vos, {\it {A robust jet
  reconstruction algorithm for high-energy lepton colliders}},  {\em Phys.
  Lett.} {\bf B750} (2015) 95--99, [\href{http://arxiv.org/abs/1404.4294}{{\tt
  arXiv:1404.4294}}].

\bibitem{Georgi:2014zwa}
H.~Georgi, {\it {A Simple Alternative to Jet-Clustering Algorithms}},
  \href{http://arxiv.org/abs/1408.1161}{{\tt arXiv:1408.1161}}.

\bibitem{Stewart:2015waa}
I.~W. Stewart, F.~J. Tackmann, J.~Thaler, C.~K. Vermilion, and T.~F. Wilkason,
  {\it {XCone: N-jettiness as an Exclusive Cone Jet Algorithm}},  {\em JHEP}
  {\bf 11} (2015) 072, [\href{http://arxiv.org/abs/1508.01516}{{\tt
  arXiv:1508.01516}}].

\end{thebibliography}\endgroup
